\documentclass{emulateapj}
\usepackage{epsfig}
\usepackage{amsmath}
\usepackage{longtable}
\usepackage{graphicx}
\usepackage{amssymb}

\usepackage{amsfonts}

\def\lcdm{$\Lambda$CDM}
\def\msun{\mbox{$M_\odot$}}
\def\ms{\mbox{$M_{\rm s}$}}
\def\mh{\mbox{$M_{\rm h}$}}   

\def\apjl {ApJL}
\def\apjs {ApJS}

\def\aap {A\&A}

\def\mstar {h^{-2}~\rm M_{\odot}} 
\def\mhalo {h^{-1}~\rm M_{\odot}} 

\shorttitle{Central galaxies in different environments}
\shortauthors{Lacerna et al.}

\begin{document}

\title
  {Central galaxies in different environments: do they have similar properties?}


\author{I. Lacerna, A. Rodr\'iguez-Puebla$^{1}$, V. Avila-Reese and H. M. Hern\'andez-Toledo}
\affil{Instituto de Astronom\'ia, Universidad Nacional Aut\'onoma de M\'exico, A. P. 70-264, 04510, M\'exico, D.F., M\'exico}


\altaffiltext{1}{Also Center for Astronomy and Astrophysics, Shanghai Jiao Tong University, Shanghai 200240, China.}

\newtheorem{theorem}{Theorem}[section]

\label{firstpage}


\begin{abstract}
We perform an exhaustive comparison among central galaxies from SDSS catalogs in different local environments at $0.01\le z \le0.08$. The central galaxies are separated into two categories: group centrals (host halos containing satellites) and field centrals (host halos without satellites). From the latter, we select other two subsamples: isolated centrals and bright field centrals, both with the same magnitude limit. The stellar mass (\ms) distributions of the field and group central galaxies are different, which explains why in general the field central galaxies are mainly located in the blue cloud/star forming regions, whereas the group central galaxies are strongly biased to
the red sequence/passive regions. The isolated centrals occupy the same regions as the bright field centrals since both populations have similar \ms\ distributions. At parity of \ms, the color and specific star formation rate (sSFR) distributions of the samples are similar, specially between field and group centrals. Furthermore, we find that the stellar-to-halo mass (\ms--\mh) relation of isolated galaxies does not depend on the color, sSFR and morphological type. For systems without satellites, the \ms--\mh\ relation steepens at high 
halo masses compared to group centrals, which is 
a consequence 
of assuming a one-to-one relation between group total stellar mass and halo mass.
Under the same assumption,
the scatter around the \ms--\mh\ relation of centrals 
with satellites increases with halo mass.
Our results suggest that the mass growth of central galaxies is mostly driven by the halo mass, 
with environment and mergers playing a secondary role.
\end{abstract}
\section{Introduction}

According to the current cosmological paradigm, galaxies form inside growing Cold Dark Matter (CDM) halos. 
The mass assembly of the CDM halos is hierarchical and with time many halos that were distinct (not contained
inside larger halos) become subhalos. In this way, 
galaxies evolving in the centers of 
the halos and subhalos
become part of gravitationally bounded groups, 
with a \textit{central} galaxy residing in the 
center of the host halo and \textit{satellite} galaxies residing in the orbiting subhalos.\footnote{The term `subhalos' is sometimes used for substructures containing  central or satellite galaxies. For example, in numerical simulations the main subhalo contains a central galaxy in the halo. However, in this work we refer to subhalos as the substructures that are contained inside a distinct halo and that host only satellite galaxies.}
Therefore, all observed local galaxies can be classified 
as centrals and satellites. One expects that
the evolution of centrals is mainly driven by internal processes while the evolution of satellites is likely 
affected by 
environmental effects of the host halo, both at the dynamical (tidal stripping) and 
hydrodynamical (starvation, ram pressure, harassment, induced star formation, etc.) levels.  

In this paper, we will focus on the properties of central galaxies of halo-based groups. It is important 
to highlight the differences between the classical observational definitions of groups and clusters in 
astronomy and the one of halo-based groups that will be used in this paper. The latter is a more general definition
since it includes groups and clusters but also systems like the Milky Way and their satellites. The halo-based 
group concept uses the halo radius (typically the virial one) as the criterion for defining the membership of the 
central and satellites galaxies to the group \citep[see for more details][]{Yang+2005, Yang+2007}. 
Central galaxies in distinct halos may have many, few or no 
satellites above a given limit in luminosity or stellar mass. 
The galaxy group configuration changes actually with time, typically in the direction of the central galaxy growing by 
merging satellites (galactic cannibalism) and by acquiring new satellites.  The observed configuration of the group,  
e.g., the absence of satellites or the gap between the masses (luminosities) of the central and the most 
massive satellite, may reveal its dynamical degree of evolution; in the context of the \lcdm\ cosmology, the most
massive halo-based groups (clusters of galaxies) are on average dynamically younger than the 
less massive ones.   
There emerges a natural question: is the group configuration related to the properties of the 
central galaxy? Are the masses, colors and star formation rates (SFR's) of centrals without satellites different
from those with satellites? Among the latter, are there differences between those with small and large gaps in mass?

Among central galaxies, those 
in extremely isolated environments are the ones 
whose evolution is
expected to be less affected 
by external physical processes. In this sense, isolated galaxies are considered as optimal objects for 
constraining the internal physical processes of modeled and simulated galaxies (e.g., dynamical assembly of 
disks and spheroids, star formation and its feedback, and AGN feedback). 
From the observational point of view, it is not an easy task to define optimal isolation criteria and apply them
to large galaxy samples. An early attempt to construct such a sample was carried out by \citet{Karachentseva+1973},
who
compiled the Catalogue of Isolated Galaxies in the northern hemisphere (CIG) 
that consists of 1050 galaxies found by visual
inspection of the Palomar photographic plates, with isolation criteria based on apparent diameters, projected distances, 
and definite size ratios between 
candidate isolated galaxies and potential perturber neighbors.
Besides the CIG and its 
recent refinements (see below), 
other catalogs with different isolation criteria 
have been compiled and used as control samples in studies of galaxies 
in different environments
\citep[][among others]{Marquez+1999, Aars+2001, Varela+2004, Verley+2007}.

The CIG is a magnitude-limited sample that is approximately complete up to $mZW \sim$ 15.5 (blue magnitudes) with a well defined selection function. 
After the extensive homogeneous spectroscopic and imaging data releases from digitized sky surveys,
several works improved the observational properties of the objects in the CIG catalog \citep[e.g.,][]{Verdes-Montenegro+2005,
Hernandez-Toledo+2007, Hernandez-Toledo+2008}.
In \citet{Hernandez-Toledo+2010} the isolation criterion
has been refined to include information on the relative recessional velocities of the galaxies and their neighbors (a 3D selection instead of a 2D-isolation criterion)
as an attempt to avoid non-physical (projected) companions. 
These authors 
have applied the refined CIG isolation criteria to the Sloan Digital Sky Survey (SDSS) DR5 \citep[][]{DR5+2007} and found
1520 isolated galaxies above 15.2  
$r$-band apparent magnitudes
in the sample dubbed as UNAM-KIAS.

We analyze the properties of central galaxies from the 
\citet[][hereafter Y07]{Yang+2007}  
halo-based group catalog
and from the isolated UNAM-KIAS catalog. 
We can separate the Y07 sample into centrals
with satellites
and those with not detected satellites (the only galaxy in the halo is the
central one).  
In addition, we select those centrals which correspond to 
very isolated galaxies in the UNAM-KIAS catalog.
It is known that satellite galaxies show properties 
that are different 
when compared to central galaxies
mainly due to 
environmental effects of the host halo
\citep[e.g.,][and more reference therein]{Weinmann+2006,vandenBosch+2008,Weinmann+2009,Pasquali+2010,Peng+2010,Peng+2012,Woo+2013,Wetzel+2013}.
Our aim is to explore 
the differences in properties
among centrals 
with a massive satellite, with smaller satellite(s), without satellites, and in very isolated environments. This exercise
is important for evaluating the use of central and isolated galaxies as control objects
in studies aimed to constrain galaxy evolution driven by internal physical processes. 
  
The outline of this work is as follows. In Section \ref{sec_data}, we present the data set of central galaxies studied in this work. We compare observational properties such as color and specific SFR among the different categories of central galaxies in Section \ref{secObs}. 
In Section \ref{secMh_conn}, we study the stellar-to-halo mass relation of central galaxies, with an emphasis on how this relation depends on the properties of very isolated galaxies.
We discuss our results in Section \ref{reasons}. Finally, in Section \ref{secConcl}, we give our conclusions.


\section{Data}
\label{sec_data}

The aim of this paper is to analyze the properties 
of central galaxies in different environments. For this purpose, we use the general 
galaxy group catalog constructed by Y07 and the 
catalog for isolated galaxies reported in \citet[][]{Hernandez-Toledo+2010}.  
We select the central galaxy as the most massive object 
within the halo. 
This seems to be a reasonable assumption for halos less massive than $\sim 2\times 10^{13}$ \msun (which is our general case; see the bottom panel of Fig. \ref{frac_Ms-Mh} below). According to 
a study carried out by \citet{Skibba+2011}, the fraction of most massive galaxies which are not the
centrals in these halos is $\sim 0.25$. This fraction increases to $\sim 0.4$ for the most massive halos.


The first catalog is extracted from a more general galaxy group sample constructed by Y07 
from the New York University Value-Added Galaxy Catalog
\citep[NYU-VAGC;][]{Blanton+2005}, which is based on 
SDSS DR4 \citep[][]{DR4+2006}. 
By using a halo-based group finder algorithm, 
Y07 \citep[see also][]{Yang+2008,Yang+2009,Yang+2012} associated dynamically to the galaxies a dark matter halo.
This group system may contain one or more galaxies 
and extends up to the virial radius of the given halo.  
The idea behind the group finder consists of an iterative 
procedure that uses average mass-to-light ratios of groups, based on the total luminosity of all group members 
down to some luminosity, to assign a tentative mass to each group. Then the virial radius associated to this 
mass is used to recalculate the group membership, repeating this process until convergence is reached. 
Y07 have tested this method by constructing mock catalogs  based on the 
SDSS and found that  $80\%$ have a completeness greater than 0.6, 
while $85\%$ have a contamination by interlopers lower than 0.5. 
The full sample consists of 369 447 galaxies with redshifts in the range 0.01 $\leq$ $z$ $\leq$ 0.2, where the $\sim$80\% 
of them are central galaxies. Hereafter we refer to central galaxies in halos without satellites as \textit{field centrals}, $N$ = 1, 
and those central galaxies in halos that host satellites as \textit{group centrals}, $N$ $>$ 1, where $N$ is the total number 
of galaxies within a halo.

Since fiber collisions could introduce some systematic error in the Y07 group identification algorithm, it is important to study its impact. To that end, \citet{Yang+2009} 
divided the group catalog into two samples: one that uses galaxies with known redshifts and another that includes galaxies which lack redshifts due to fiber collisions. They found that the conditional stellar mass function for the fiber collision-corrected sample 
has a higher amplitude than 
that in the non-corrected case,
particularly for low-mass halos. Nevertheless, the differences are marginal and well within the error bars. Therefore, we conclude that fiber collisions in the Y07 sample are not a source of systematic errors that could affect our conclusions.

The second catalog comes from the UNAM-KIAS collaboration 
\citep[][]{Hernandez-Toledo+2010}.
They identified a total of 1520 isolated galaxies  
from the SDSS DR5 using 
an improved method based on the 2D-criterion 
of isolation proposed by \citet{Karachentseva+1973}. 
In addition to the condition that the projected separation from 
a neighbor across the line of sight is greater than 100 times the seeing-corrected Petrosian radius of the neighbor galaxy,
the new method takes into account that the radial velocity difference with respect to a neighbor is greater than 1000 km s$^{-1}$  
to mimic a 3D-criterion. The velocity information on galaxies in the radial direction is used 
to tackle projection effects as much as possible.
In cases when a candidate (isolated) galaxy has close neighbors (i.e., it does not satisfy the 3D-criterion), this galaxy can be considered as isolated if 
the extinction-corrected apparent Petrosian $r$-band  magnitude difference between the candidate galaxy and any neighbor is 2.5 mag ($\Delta m_{r} \ge 2.5$). This condition allows an isolated galaxy to have close (and fainter) neighbors but rejects relevant perturbers.
As the magnitude limit of SDSS is $m_{r} \sim$ 17.7, only galaxies brighter than $m_{r}$ = 15.2 were used to select isolated galaxies to take into account the magnitude difference of $\Delta m_{r} \ge 2.5$.
Most of the isolated galaxies have a redshift distribution between $z \sim 0$ and $z \sim 0.08$, with a mean redshift of $\big<z\big>$ = 0.032. 
After using an image processing scheme,
the catalog contains the information on several structural parameters, including the morphological parameter $T$.
For more details regarding the UNAM-KIAS catalog see 
\citet{Hernandez-Toledo+2010}.

In general isolated galaxies are central objects within their host halos,
although a small fraction of them can be satellite galaxies residing in the outskirts of parent halos \citep{Hirschmann+2013}.
Formally, we define our sample of isolated central galaxies as the intersection of the UNAM-KIAS catalog 
with the one of field centrals ($N=1$) in the Y07 catalog.
There are 
1046 isolated galaxies from UNAM-KIAS in Y07. 
The rest of isolated galaxies were not identified because 
of the different redshift distributions of both catalogs (see 
Section \ref{volume}) or because they belong to different data releases.
We excluded from our study 4 out of 1046 isolated galaxies which are classified as satellites in Y07 according to their stellar mass
(one of them is the central object according to their luminosity, 
other two galaxies are located in the outskirts of relatively massive groups and the remaining isolated satellite galaxy resides within a halo that suffers strong survey edge effects).
We found other 220 isolated galaxies which are centrals in groups that host satellites 
($N>1$). Recall that the isolation criterion requires not to have companions more luminous than 2.5 mag the apparent magnitude of the primary; galaxies with fainter companions (i.e., satellites) than this threshold are considered as isolated objects. We find that 85\% of the 220 
isolated galaxies in halos with satellite(s) according to Y07, have their most massive satellite 
below 0.1 times the mass of the central. Therefore, these satellites are likely 
considered as not relevant perturbers by the isolation criterion of the UNAM-KIAS collaboration.
On the other hand, note that the group finder algorithm of Y07 suffers a contamination of around 10\% when classifying 
centrals and satellites.
Rigorously, we decided to exclude from our analysis those 220 isolated central galaxies that appear with satellite(s) in the Y07 catalog. 
There remain then 822 isolated galaxies that are centrals in halos with no satellites ($N$ = 1) in Y07. 
These galaxies conform our robust \textit{isolated central} galaxy sample.

Since the sample of isolated central galaxies is brighter than $m_{r}$ = 15.2, 
we impose the same apparent magnitude limit 
on the field central galaxy sample 
in order not to introduce a selection
bias when making comparisons among them.
We refer to this subsample as the \textit{bright field centrals}. 
Recall that the central galaxies without satellites can be located in a wide range of environments, including the very isolated one. The isolated galaxies are actually those (bright) field centrals obeying extreme isolation criteria. Based on these criteria, we conclude that the subsample of isolated galaxies differs in its local environment from the bulk of the bright field centrals. 
Our subsample of isolated centrals is in fact an extreme of the distribution of all field central galaxies regarding environment.

\subsection{Volume and mass limits}
\label{volume}

Since the UNAM-KIAS catalog reaches mainly
out to $z=0.08$ and Y07 catalog ranges from $z=0.01$ out to $z=0.2$, we select those galaxies that are located within 
the same volume in redshift space in order to have a fair comparison between central galaxies from both catalogs. 
Therefore, throughout the text we use the redshift range 0.01 $\leq$ $z$ $\leq$ 0.08.
As explained below, we characterize the galaxies from all of our samples by their stellar mass, \ms.
In a magnitude limited sample, the minimum detected \ms\ depends on the redshift and on the stellar
mass-to-luminosity ratio; the latter depends on the stellar populations and therefore on galaxy colors.
For the SDSS sample and its magnitude limit, \citet[][see also \citealp{Yang+2009}]{vandenBosch+2008} 
have calculated the stellar mass limit at each $z$ above which the sample is complete, i.e., galaxies are potentially 
observable above this mass:
\begin{eqnarray} 
\lefteqn{ \textrm{log}[M_{s,lim}/\mstar] =} \nonumber\\
&  & {} \frac{4.852 + 2.246 \textrm{ log}(d_L) + 1.123 \textrm{ log}(1+z) - 1.186z}{1-0.067z}.
\label{eq_Mslim}
\end{eqnarray} 
We adopt this limit.

{\it After imposing the above mentioned  volume and mass limits to our samples, there remain 
822 isolated, 10 708 group, 40 551 field and 4358 bright field central galaxies}. These are the
main samples to be used in this paper. A summary of their properties is given in Table \ref{tabla_samples}.

\begin{table*}
  \centering
\caption{General properties of the central galaxy samples}
\begin{tabular}{c c c c c c c c}
\hline
\hline
sample           & N     & source catalog & $z$-range                   & m$_r$ limit   & n      &  n$_{M_h}$\\  
(1)              & (2)   & (3)            & (4)                      & (5)           & (6)    & (7) \\
\hline
group            & $>$ 1 &  Y07           &  0.01 $\leq z \leq$ 0.08 &  $<$ 17.77    & 10708  &   9271 \\
field            & = 1   &  Y07           &  0.01 $\leq z \leq$ 0.08 &  $<$ 17.77    & 40551  &  28043 \\
bright field$^a$ & = 1   &  Y07           &  0.01 $\leq z \leq$ 0.08 &  $<$ 15.2     &  4358  &   3420 \\
isolated$^b$     & = 1   &  UNAM-KIAS     &  0.01 $\leq z \leq$ 0.08 &  $<$ 15.2     &  822   &    663 \\                    
\hline
\end{tabular}

\tablecomments{Column 1: sample name. Column 2: galaxy number within the host halo. Column 3: source catalog. Column 4: redshift range. Column 5: extinction-corrected apparent Petrosian $r$-band magnitude limit. Column 6: galaxy number of each sample. Column 7: number of galaxies with halo mass estimates.}

\tablenotetext{1}{Subsample of the field central galaxies with $m_r < 15.2$.}
\tablenotetext{2}{Isolated galaxies identified in the bright field sample.}
\label{tabla_samples}
\end{table*}

\subsection{Stellar and gas masses, colors and sSFR}
\label{secMs}

In Section \ref{secObs} we will compare observational properties such as color and specific star formation 
rate (sSFR) in bins of stellar mass (\ms) for our samples of central galaxies. Details of these properties are given in this Section.

Stellar masses for all the central galaxies are taken from Y07, who used the relation between the stellar 
mass-to-light ratio and color of 
\citet{Bell+2003}.  We note that in the processed MPA-JHU DR7 catalog,\footnote{Available at http://www.mpa-garching.mpg.de/SDSS/DR7/}
stellar masses are also provided for each galaxy in the Y07 group catalog, but calculated in this case through 
the spectral energy distribution fitting. While this method offers a more constrained estimate for the
stellar masses than the \citet{Bell+2003} method, the differences
between the two mass estimates are small and not systematical \citep[see for a comparison e.g.,][]{Dutton+2011}.

For the isolated central galaxies, we are also interested in the study of how their gas content 
compares with respect to observations.
We include the information on the H I line magnitude (corrected for self-absorption) from the HyperLeda database\footnote{http://leda.univ-lyon1.fr/} for the sample of isolated galaxies in order to estimate their gas mass $M_{gas}$ content. The neutral hydrogen line magnitude in 21 cm, $m_{21}$, is converted into H I mass $M_{HI}$ by using 
\begin{eqnarray} 
\frac{M_{H I}}{\msun} = 2.36 \times 10^5 \textrm{ } d_L^2 \textrm{ } 10^{(17.4 - m_{21})/2.5}  \textrm{ ,} 
\label{eq_MHI}
\end{eqnarray} 
where $d_L$ is the luminous distance in Mpc 
\citep[][]{RH1994} 
which is calculated under the $\Lambda$CDM cosmology with $\Omega_m = 0.27$, $\Omega_{\Lambda} = 0.73$. 
The gas mass can be calculated by using
a simple correction for helium and metals, $M_{gas}$ = 1.4$M_{HI}$, or, more accurately, by adding 
a correction for 
the molecular hydrogen H$_2$, $M_{gas}$ = 1.4$M_{HI}$(1 + $M_{H_2}/M_{HI}$). The ratio  
$M_{H_2}/M_{HI}$ depends on the morphological parameter $T$, which is available in the UNAM-KIAS catalog.
This ratio is given by $M_{H_2}/M_{HI}$ = 3.7 - 0.8$T$ + 0.043$T^2$ 
\citep[][]{McGaugh_deBlok1997}.
As noted by \citet{Avila-Reese+2008}, 
this empirical relation is valid for late-type spiral galaxies 
($T$ $\geq$ 2) since it can overestimate the gas mass for galaxies with earlier morphologies. For galaxies with $T$ $<$ 2, we use their assumption $M_{H_2}/M_{HI}$ = 2.3.

In Section \ref{secObs} we use two widely separated bands for the color, namely $g - i$.  
The $g - i$ color is measured using $modelMag$ magnitudes of SDSS corrected for extinction (\verb|dered| parameter in CasJobs).\footnote{See http://casjobs.sdss.org/CasJobs/} This magnitude is defined as the better of two magnitude fits; a pure de Vaucouleurs profile and a pure exponential profile. The inclination of the galaxies combined with dust 
extinction in the disk produces a systematic reddening of their colors.
In order to ameliorate this systematic effect, throughout this paper we will not use the colors 
of those galaxies with inclination angles higher than 70 degrees. 
The inclination angle $inc$ is calculated as 
$cos(inc)=b/a$, where $b/a$ is the $r$-band minor-to-major isophotal axis ratio from SDSS.

Finally, we include in our samples the sSFR, defined as the SFR divided by the stellar mass. This
quantity 
has been obtained from the MPA-JHU DR7 catalog, and it is an updated version of the estimates presented
in \citet{Brinchmann+2004} by using a spectral synthesis fitting model.

\subsection{Halo mass}
\label{secMh}

The Y07 group catalog includes by construction the halo (virial) mass, \mh, of the central galaxies down to some luminosity.
This kind of halo-based group finders have emerged as a powerful method for estimating group halo masses, 
even when there is only one galaxy in the group halo.  By using mock catalogs that resemble observations, it has 
been shown that the group masses estimated with this method can recover successfully the true halo mass,  in a statistical sense,
with no significant systematics \citep{Yang+2008}. It was also shown that most of the scatter in the relation between the 
true and assigned halo masses in the mock catalog is owed to the intrinsic scatter; the fact that the group finder is 
not perfect, i.e., suffers from interlopers and incompleteness, and that it is necessary to correct the characteristic luminosity
for members that do not make the magnitude limit of the survey, only adds a relatively small contribution to the total 
scatter. 

The Y07  
catalog contains the halo membership, the identification of central and satellite galaxies in the halo, and \mh\ based on either the  
characteristic stellar mass or the characteristic luminosity in the group, among other halo properties. We use the halo mass 
based on the characteristic stellar mass. The characteristic stellar mass 
(luminosity) is defined as the sum of the stellar mass (luminosity) of all the galaxies in the halo with $^{0.1}M_r$ - 5 log$(h) \leq$ -19.5, 
where $^{0.1}M_r$ is the absolute magnitude in the $r$-band with
$K$-correction and evolution-correction at $z$ = 0.1. 
It is also considered the completeness of the survey 
at the redshift of each of these galaxies, as well as a correction factor for the apparent magnitude limit of the spectroscopic survey.
They assume a one-to-one relation between the characteristic stellar mass (luminosity) and \mh\ by matching their rank 
orders for a given volume and a given halo mass function.

The obtained \ms--\mh\ relation is robust: by using it in mock catalogs, the average relations of halo occupation statistics 
are recovered (Y07).
This approach has advantages in a statistical sense over other methods 
used for estimating group (halo) masses, such as the one based on velocity dispersions, that needs a significant number of 
members to calculate dynamical masses, or the gravitational lensing or X-ray emission method, that need data of high quality 
which are applied only to massive systems.
However, for groups/single-galaxy systems which are not complete in characteristic stellar mass (luminosity), the halo mass 
estimates under the assumption of the one-to-one relation mentioned above are already not reliable; thus, the central galaxies 
fainter than the magnitude limit do not have halo mass estimates. This is why the halo mass limit in the Y07 catalog 
is $10^{11.6} \mhalo$.  
Nonetheless, we note that due to the method used in Y07 the estimated halo masses are not measurements of the true halo masses.
In addition, groups with strong survey edge effects do not have halo mass estimates  
regardless the luminosity of their members. The latter affects only 1.6\% of all Y07 groups.
The number of central galaxies with host halo masses is smaller than in the original catalog. For the samples used here, those with estimated halo masses are reduced in number to
{\it 663 isolated, 9271 group, 28 043 field and 3420 bright field central galaxies} (see Table \ref{tabla_samples}). 
These samples will be used in Section \ref{secMh_conn}.


\section{Observational properties}
\label{secObs}

\begin{figure}
\begin{center}
\leavevmode \epsfysize=8.5cm \epsfbox{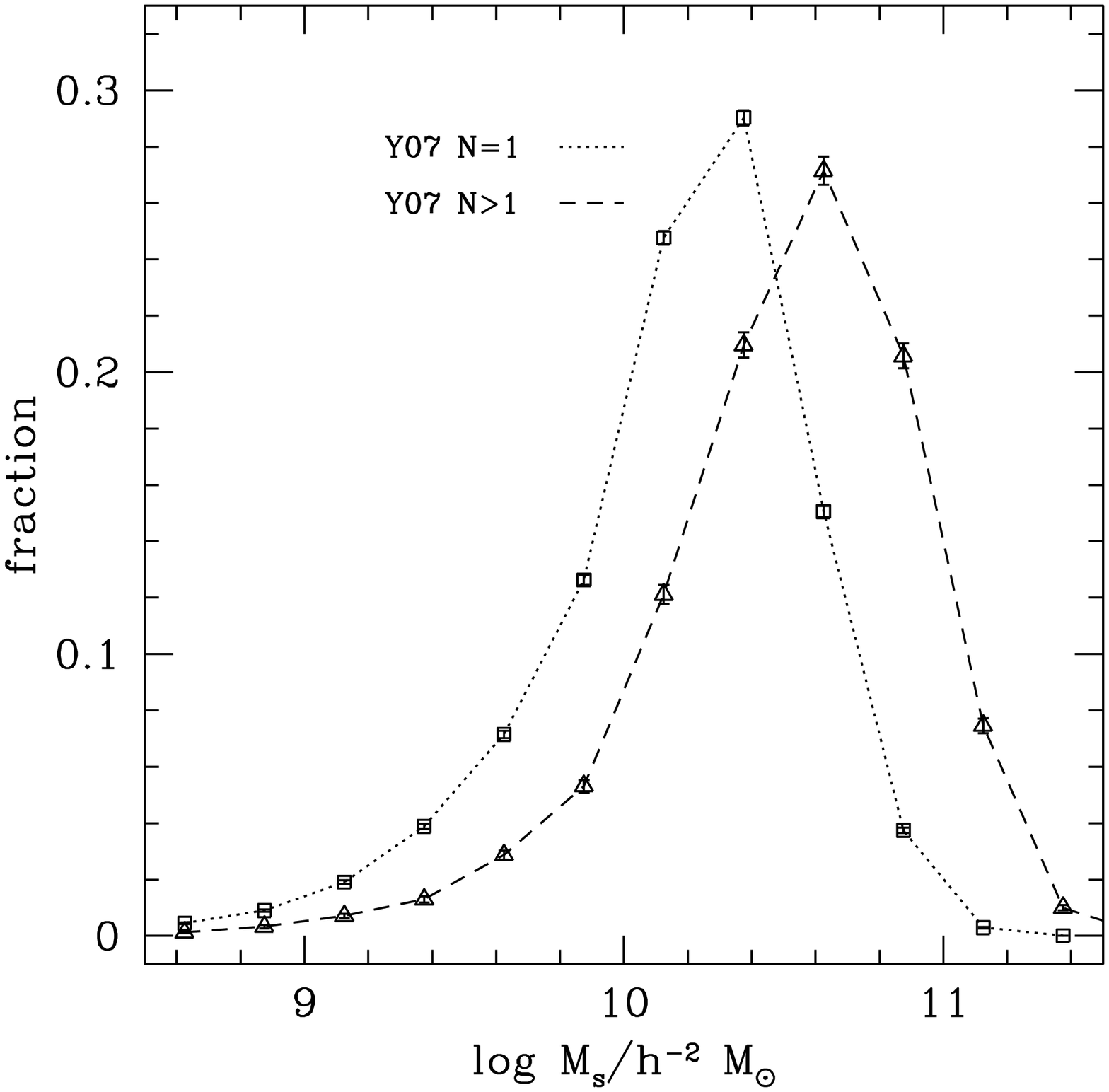} 
\leavevmode \epsfysize=8.5cm \epsfbox{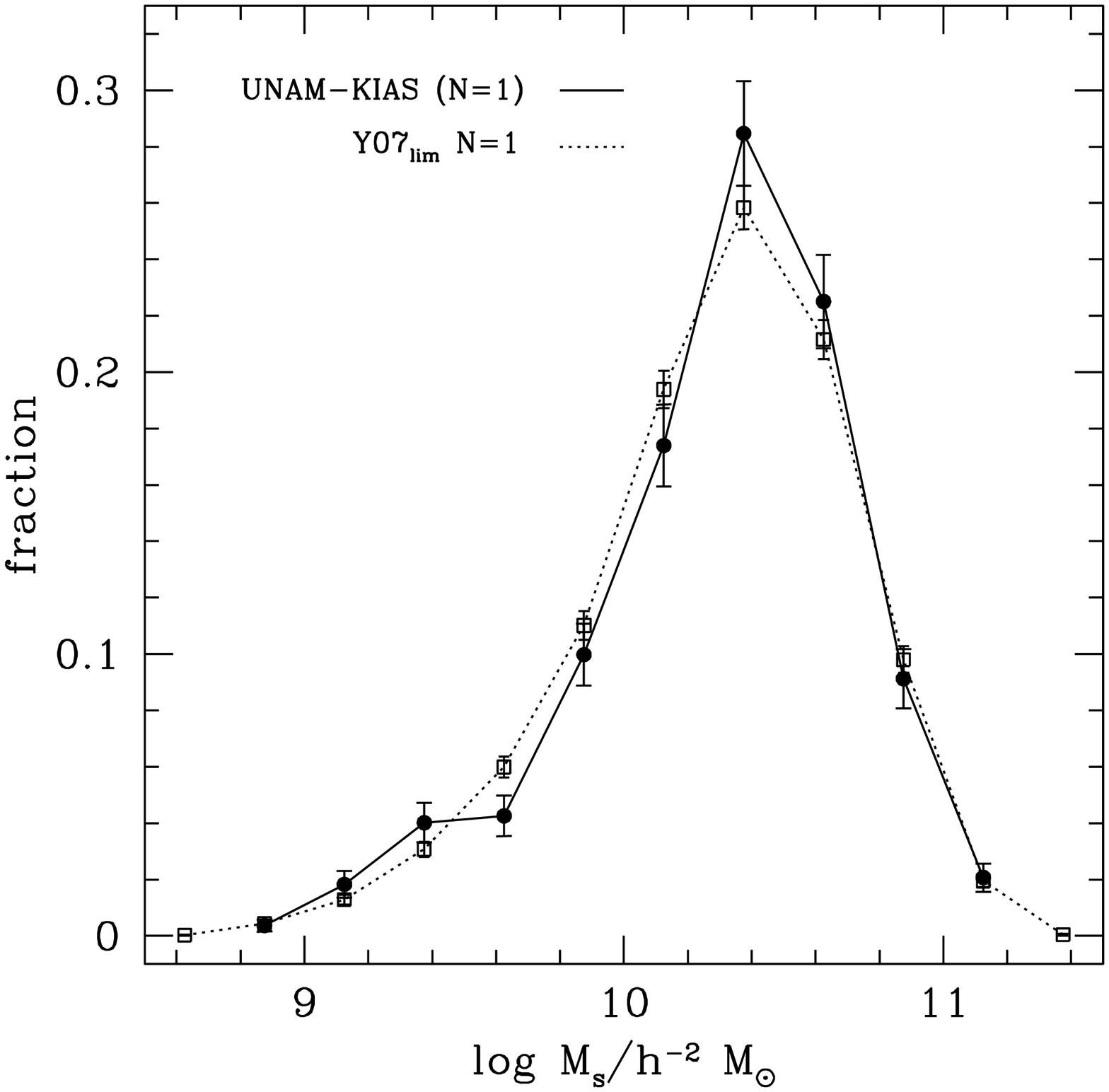} 
\caption 
{
Relative fraction of central galaxies as a function of stellar mass. The sum of the points for each category is equal to unity, therefore each line may be considered as a probability distribution. The error bars correspond to the Poisson error.
Top panel: field and group galaxies (open squares and open triangles connected by dotted and dashed lines, respectively).  
At high masses, the fraction of group centrals dominates with respect to that of field central galaxies.
Bottom panel: isolated and bright field galaxies
(filled circles and open squares connected by solid and dotted lines, respectively). Both distributions are similar within the errors.
}
\label{frac_Ms}
\end{center}
\end{figure}

\begin{figure*}
\begin{center}
\leavevmode \epsfysize=5.9cm \epsfbox{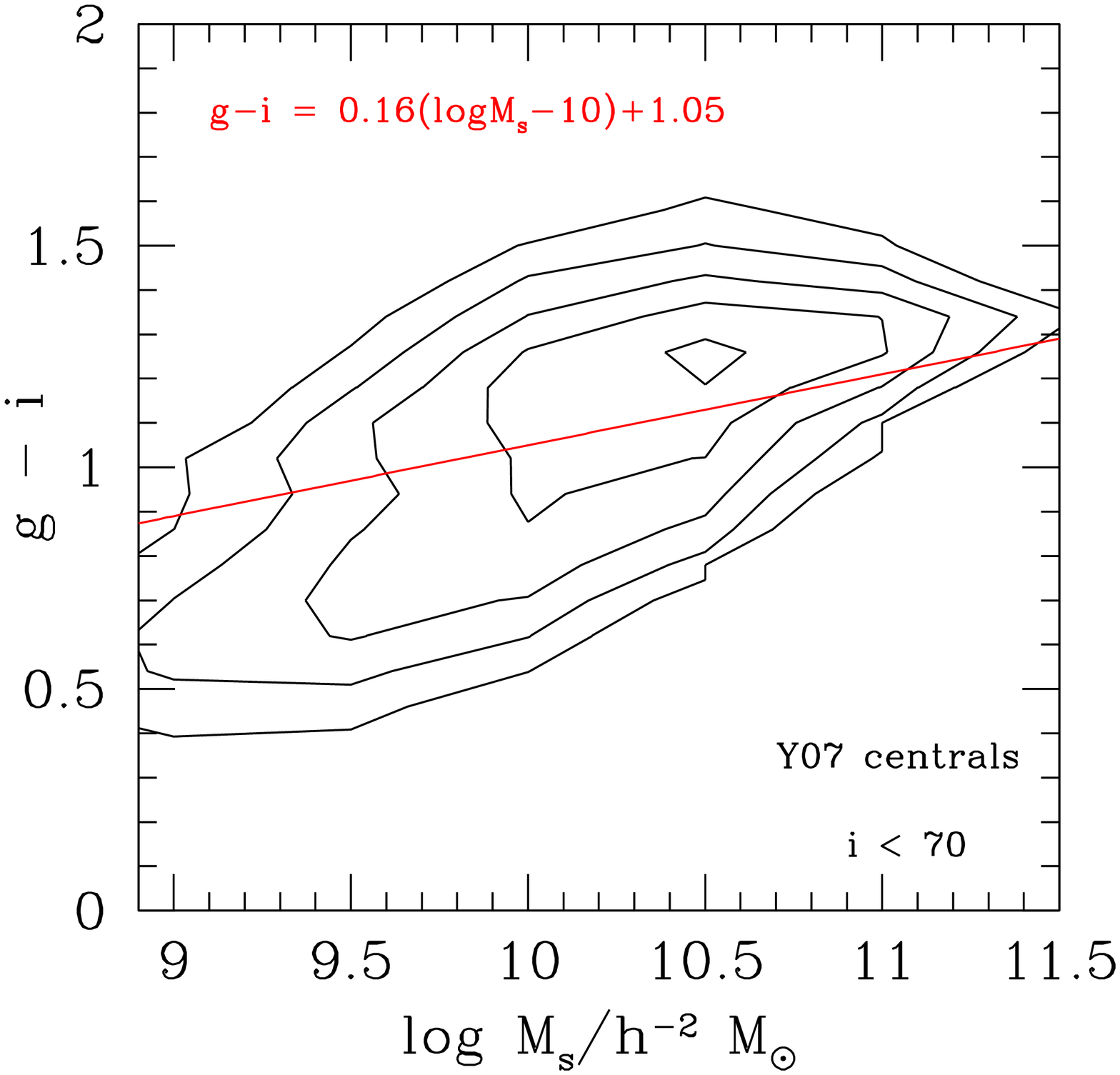} 
\epsfysize=5.9cm \epsfbox{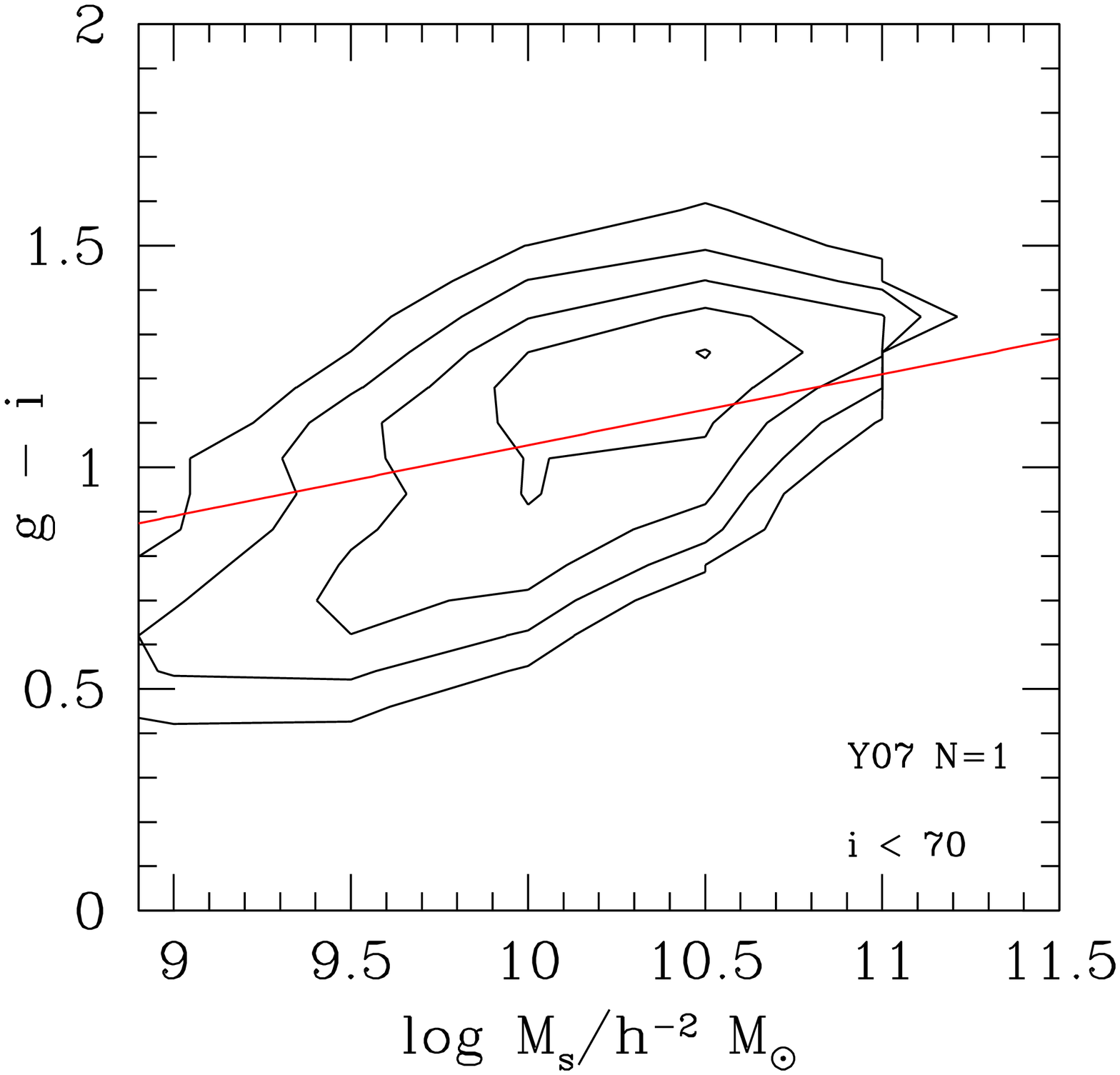} 
\epsfysize=5.9cm \epsfbox{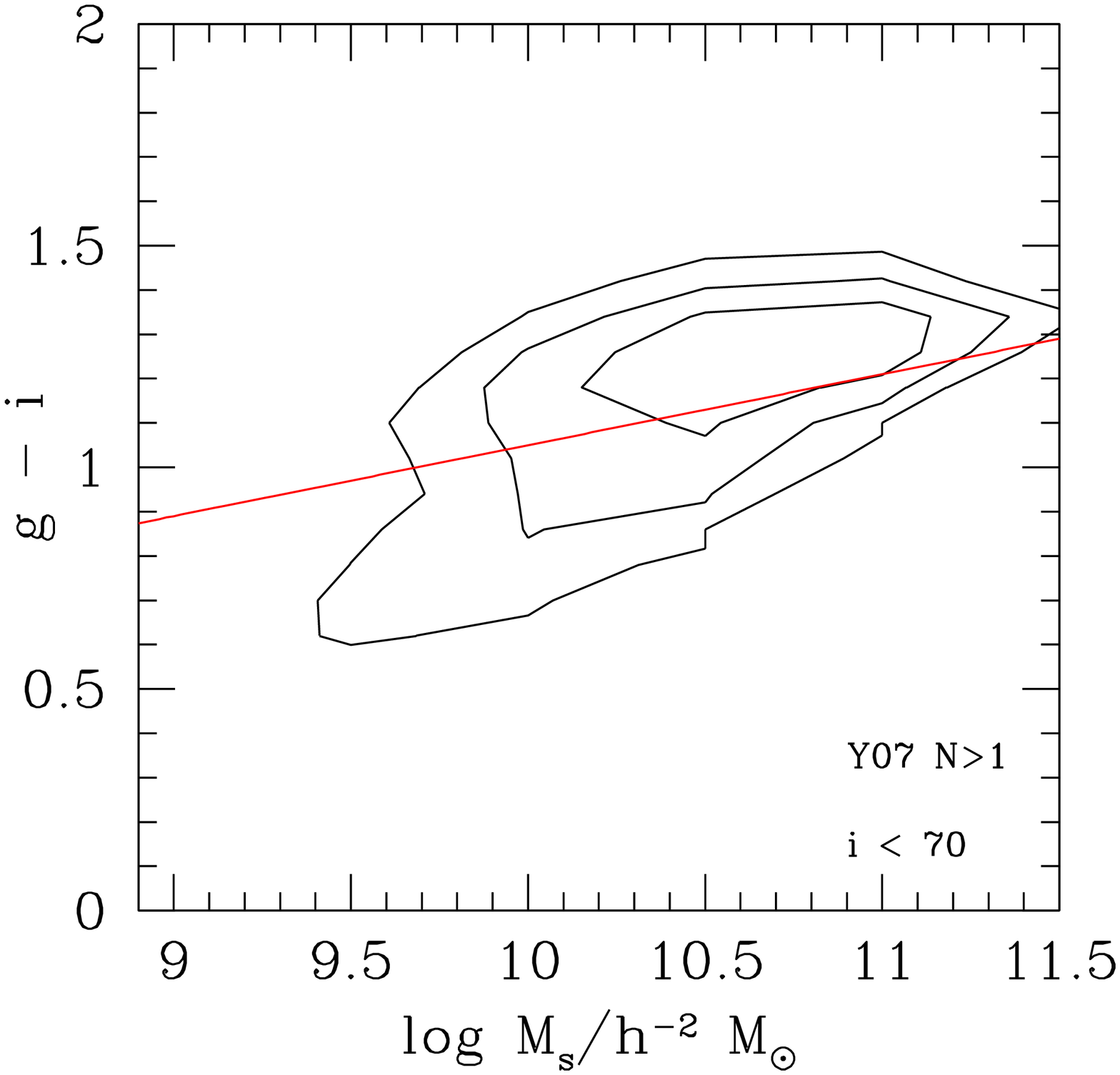}
\epsfysize=5.9cm \epsfbox{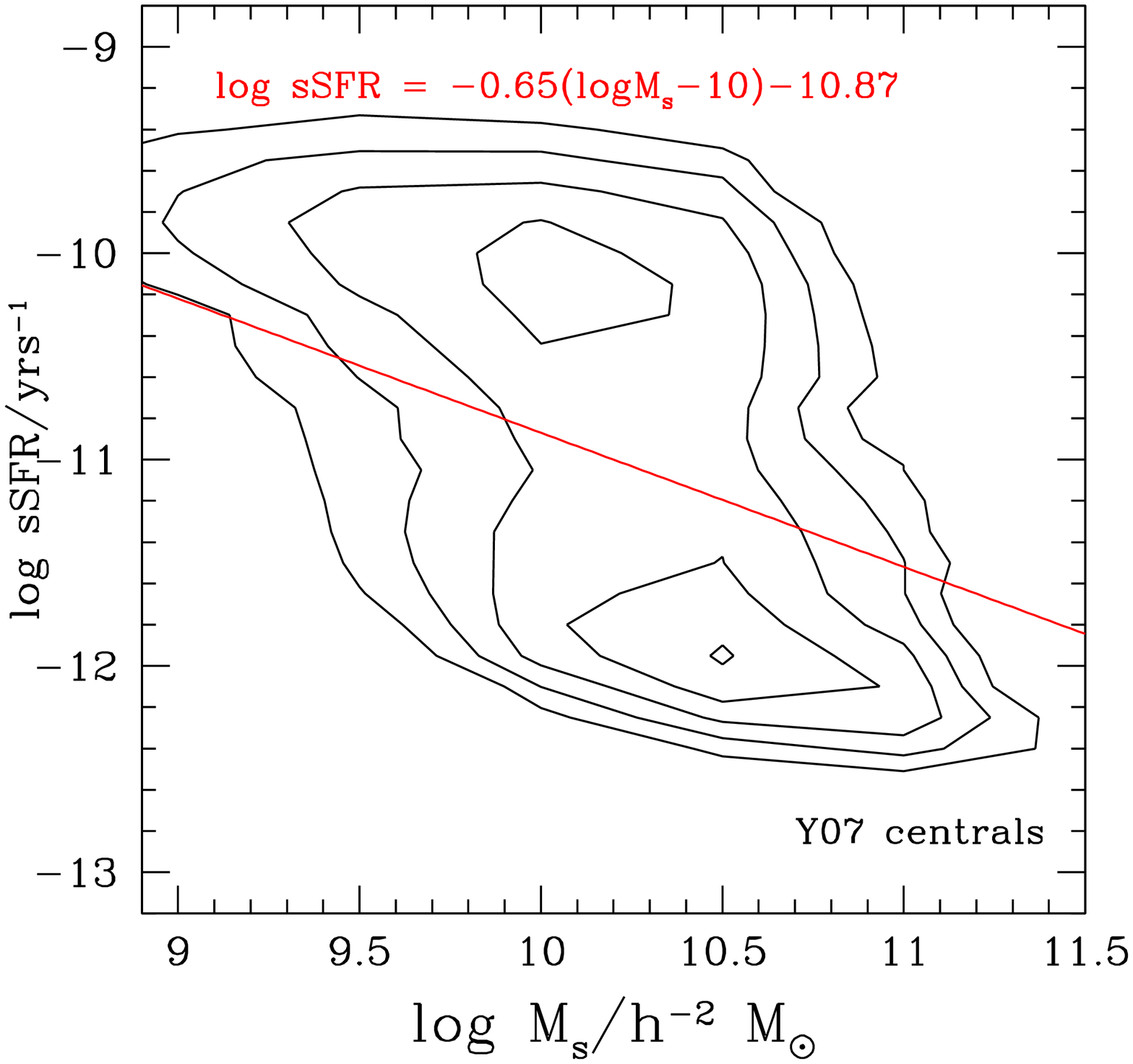} 
\epsfysize=5.9cm \epsfbox{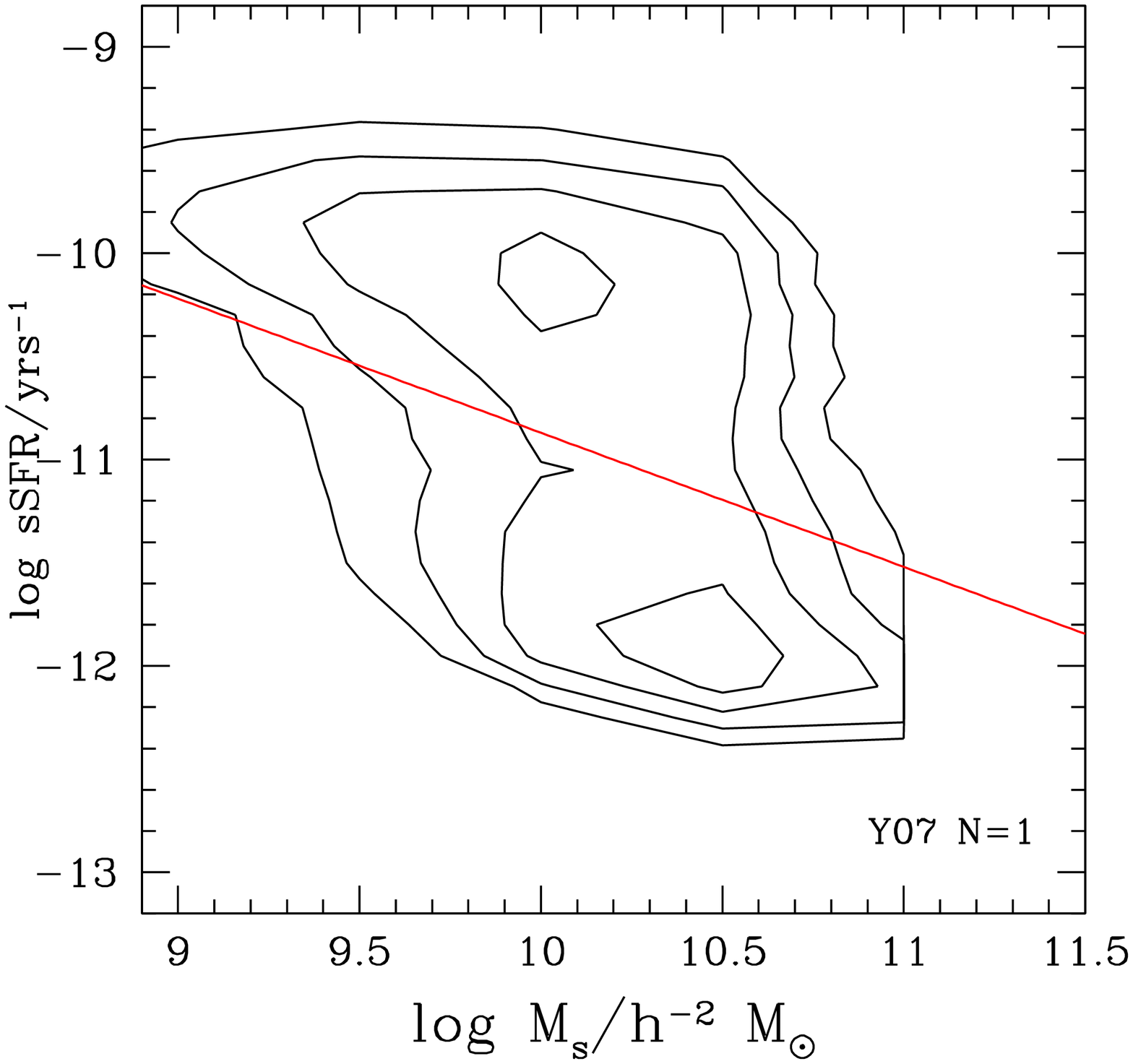} 
\epsfysize=5.9cm \epsfbox{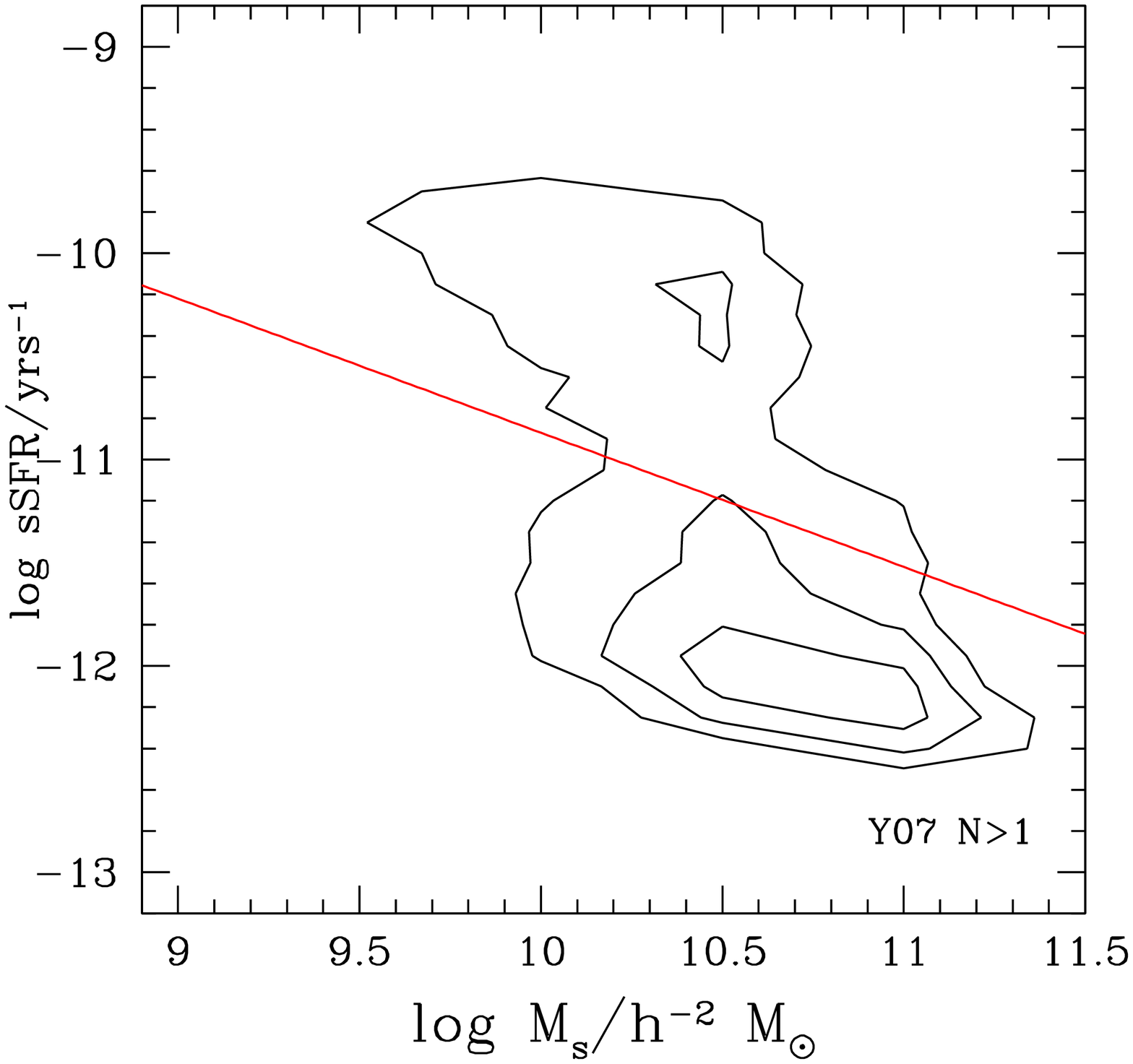}  
\caption{
Color ($g-i$, top panels) and specific star formation rate (bottom panels) as a function of stellar mass. 
All the central galaxies 
between 0.01 $\leq$ $z$ $\leq$ 0.08
are shown in contours  
in the left column. They are split in field ($N$ = 1, middle column) and group ($N >$ 1, right column) central galaxies. 
Concerning colors, we do not include objects with inclinations above 70 degrees. The red lines in the 
top and bottom panels show equations (\ref{eq_gi}) and (\ref{eq_sSFR}) to separate red/blue and passive/active 
galaxies, respectively.
}
\label{CMD_sSFR}
\end{center}
\end{figure*}

As mentioned above, we define four categories of central galaxies (see Table \ref{tabla_samples}). The 
first two categories are the \textit{group centrals} (host halos with $N>1$) 
and \textit{field centrals} (host halos with $N=1$, i.e., no satellites). The former makes the 21\% of all
central galaxies of our local (0.01 $\leq$ $z$ $\leq$ 0.08) Y07 sample, whereas the latter form the 79\%. 
The 
field centrals can inhabit all kinds of environments (e.g., around groups or clusters, 
along a filament, within a void or simply close to other galaxies), excepting those with other galaxies within their
virial radius, i.e., groups and clusters of galaxies. 
Thus, for field central galaxies
it is also possible that their near environment might be playing an important role in shaping
their properties.
In order to select field centrals in extreme isolated environments
we use the \textit{very isolated central} sample (taken from the UNAM-KIAS catalog and identified in the Y07 catalog).
Since the isolated sample has a brighter magnitude limit than the field sample, we avoid  
a selection bias when comparing both samples by using the \textit{bright field central} subsample, i.e., 
those field galaxies selected with the same $m_r$ limit that the UNAM-KIAS catalog (Section \ref{sec_data}).
The isolated galaxies
make the 19\% of the bright field sample. 
The aim of this work is to study and compare 
the properties of central galaxies selected from the group, field, and isolated samples.

Our samples are not complete in the sense of observing the full distribution of galaxies above a given (small) mass 
(e.g., see the drop in the fraction of low-mass galaxies in Fig. \ref{frac_Ms}). In fact, we will not deal with stellar 
mass functions, where corrections like the $V/V_{max}$ should be introduced. We only compare samples 
in the same redshift ranges that satisfy the stellar mass completeness condition provided by 
\citet[][eq. \ref{eq_Mslim}]{vandenBosch+2008}.

The top panel of Fig. \ref{frac_Ms}
shows the relative fraction of field and group central galaxies as a function of stellar mass (dotted 
and dashed lines, respectively).  Each distribution 
is normalized and can be considered as a probability distribution.
Since both samples are from the same parent 
volume-limited sample (0.01 $\leq$ $z$ $\leq$ 0.08), the comparison among them in Fig. \ref{frac_Ms} is fair in the 
sense that 
any incompleteness due to not observed galaxies is expected to be the same for them.
The differences among the samples of field and group central galaxies in the stellar mass distribution is noticeable. 
In addition, we performed a chi-square test, which is useful for binned data \citep[][]{NumericalRecipesC},  
that confirms a very low probability of $P < 4 \times 10^{-4}$ that both normalized distributions are similar.
The fraction of field centrals at low masses, $M_{s} <$ 10$^{10}$ $\mstar$,
is higher than the fraction of group centrals. 
At higher stellar masses, $M_{s} >$ 4 $\times$ 10$^{10}$ $\mstar$, the relative fraction of group centrals
is higher than that of field centrals.
Around 56\% of group central galaxies ($N>1$) are found at these high masses, whereas this percentage decreases to 
19$\%$ for field centrals ($N=1$).
The top panel also shows that the distributions of the centrals peak at different masses: field centrals at 
$\sim 10.4$ and group centrals at 10.6 in log($M_{s}/\mstar$). 

In the same figure, the bottom panel shows the relative fraction of isolated and bright field central galaxies as a function of stellar mass.
There are no significant differences between both distributions within the errors; the chi-square test confirms this
with a probability $P = 0.999$.
Both distributions peak 
at log($M_{s}/\mstar$)$\sim 10.4$. 
This similarity in the mass distributions suggests that the way central galaxies without satellites ($N=1$) 
attained their final stellar mass does not depend significantly on the environment further away from
the halo virial radius, at least for these bright ($m_r<15.2$) galaxies.

Figure \ref{CMD_sSFR} shows
the $g - i$ color (top panels) and sSFR (bottom panels) vs. the stellar mass. 
The left panels include all the central galaxies
between 0.01 $\leq$ $z$ $\leq$ 0.08 distributed in iso-contours 
of number density increasing by 0.5 dex.
We observe 
that the distributions of
$g-i$ color and sSFR 
depend on the stellar mass, where the bimodality in the latter is more marked than the former.
Therefore, we can separate galaxies in red/blue and passive/active by using 
\begin{eqnarray} 
g-i = 0.16[\textrm{log($M_{s}$)}-10]+1.05  \textrm{ and}
\label{eq_gi}
\end{eqnarray} 
\begin{eqnarray} 
\textrm{log(sSFR)} = -0.65[\textrm{log($M_{s}$)}-10]-10.87  \textrm{ ,} 
\label{eq_sSFR}
\end{eqnarray} 
where \ms\ is in units of $\mstar$ and sSFR is in units of yrs$^{-1}$.\footnote{These mass-dependent 
criteria for separating galaxies into two
groups, active/passive or blue/red, have been obtained by fitting two Gaussians to the corresponding (color or sSFR) 
distributions at different mass bins (see Figs. \ref{FG_gi} and \ref{FG_sSFR} below) and choosing the $g-i$ or sSFR 
values at the given bin as the ones where both Gaussians intersect in each case.}
These equations are shown as red solid lines. 
The iso-contours in the middle and right columns correspond to the samples separated into field ($N$ = 1) and group 
($N >$ 1) central galaxies. 
As can be seen in the top panels, 
the field centrals populate both the blue cloud and red sequence regions, with a preference for the former, whereas the group centrals are strongly biased to the red sequence.
Recall that we do not consider galaxies 
with high inclination angles ($inc>70$ degrees) in order to avoid systematic effects of reddening by extinction. 
The left-bottom panel of Fig. \ref{CMD_sSFR} shows the sSFR as a function of stellar mass of all the central galaxies out to $z = 0.08$, which is similar to the 
one presented in \cite{Yang+2013} by using a central galaxy sample out to $z = 0.2$. In the 
middle-bottom and right-bottom panels
we present the same but for field and group centrals, respectively. 
We find that the distribution of field central galaxies 
is strongly bimodal, whereas
the distribution of group central galaxies
shows a higher preference for occupying the passive region. Only a small fraction of group centrals are star forming galaxies.

We thus conclude that field and group central galaxies show different distributions in the 
color--$\ms$ and sSFR--$\ms$ diagrams. Central galaxies in halos without satellites ($N$ = 1; field) seem to be bluer and more 
star forming than centrals located in halos which host satellites galaxies ($N >$ 1; group). In fact, such differences 
can be explained by their differences in the stellar mass distributions. Typically, group central galaxies have higher 
stellar masses than field central galaxies (top panel of Fig. \ref{frac_Ms}). Therefore, 
considering the overall populations,
group centrals are redder, with a significant red sequence, and 
more passive than field centrals.
We also compare the samples of isolated and bright field centrals in the color--\ms\ and sSFR--\ms\ diagrams. As expected,
both samples occupy the same regions because of their similar stellar mass distributions (bottom panel of Fig. \ref{frac_Ms}).

We have found differences among central galaxies, which can be explained mainly by the differences in their stellar mass distributions. 
In the next subsections we will compare the properties of central galaxies 
in stellar mass bins. 
We begin by comparing the $g-i$ color and the sSFR distributions between the 
samples of field and group central galaxies. Then, in order to test whether 
central galaxies are affected by more extreme environmental effects, we will repeat our analysis
between the subsamples of isolated galaxies and bright field central galaxies.

\subsection{Field and group central galaxies}

\begin{table}
  \centering
\caption{Relative fractions of blue field and blue group central galaxies
}
\begin{tabular}{c c c c c}
\hline
\hline
log $M_s$    &  $g-i$ cut   & $f_{blue}$ field & $f_{blue}$ group \\   
\hline
9.2--9.6       & 0.95     &  0.79      &  0.84 \\
9.6--10.0      & 1.02     &  0.65      &  0.67 \\
10.0--10.4     & 1.08     &  0.31      &  0.33 \\
10.4--10.8     & 1.15     &  0.18      &  0.21 \\                    
10.8--11.2     & 1.21     &  0.12      &  0.13 \\                      
\hline
\end{tabular}

\tablecomments{
Stellar mass bins as shown in Fig. \ref{FG_gi}.
The second column indicates the cut in $g-i$ to define a blue/red galaxy.
}
\label{tabla_FG_gi}
\end{table}

\begin{figure}
\leavevmode \epsfysize=8.6cm \epsfbox{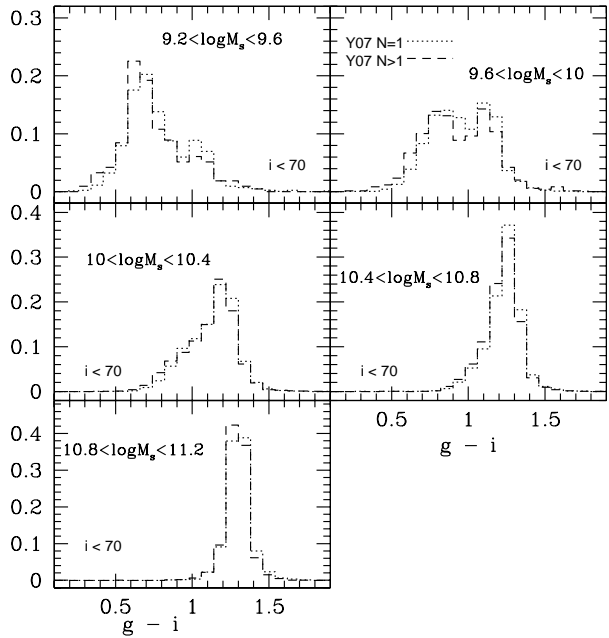}
\caption 
{Histograms of $g - i$ color for field ($N$ = 1; dotted lines) and group (N $>$ 1; dashed lines) central galaxies for 
different stellar mass bins. The stellar mass \ms\ is in units of $\mstar$.
We do not include objects with high inclination (angles over 70 degrees).
Each histogram was normalized by the total number of selected objects, so that
the sum of each distribution is equal to unity.
}
\label{FG_gi}
\end{figure}

Fig. \ref{FG_gi} shows a series of $g - i$ color histograms for field ($N$ = 1; dotted lines) and 
group (N $>$ 1; dashed lines) central galaxies in different stellar mass bins. Each distribution is 
normalized so that its sum is equal to unity. 
At a qualitative level, we see that there are no significant 
differences in the color distributions between both populations at a given stellar mass. 
The chi-square test
confirms with a very high probability ($P > 0.999$) that both normalized distributions are similar in each panel.
Table \ref{tabla_FG_gi} reports the corresponding fractions of blue central galaxies, $f_{blue}$ (the 
complement is the fraction of red centrals, $f_{red}=1-f_{blue}$) in stellar mass bins as shown in Fig. \ref{FG_gi}. 
The criterion to define blue and red galaxies is the one given in eq. (\ref{eq_gi}).
For low-mass galaxies (top  panels), most of them are blue, though
a small population of redder galaxies appears (slight bimodality).
We measure that the fraction of red group centrals is at most 5\% lower than that of red field centrals. 
At intermediate masses (middle panels), the slight bimodality disappears, 
and both populations are typically 
red. 
Similarly,
for high-mass galaxies (bottom panel), 
nearly all the field and group central galaxies are red with the 
peak around $g - i$ = 1.3.

\begin{table}
  \centering
\caption{Relative fractions of active field and active group central galaxies
}
\begin{tabular}{c c c c c}
\hline
\hline
log $M_s$    &  sSFR cut   & $f_{active}$ field & $f_{active}$ group \\   
\hline
9.2--9.6     & -10.5     &  0.82      &  0.84 \\
9.6--10.0    & -10.7     &  0.71      &  0.70 \\
10.0--10.4   & -11.0     &  0.53      &  0.52 \\
10.4--10.8   & -11.3     &  0.33      &  0.31 \\
10.8--11.2   & -11.5     &  0.14      &  0.11 \\                                            
\hline
\end{tabular}

\tablecomments{\ms\ bins as shown in Fig. \ref{FG_sSFR}. The second column 
is the cut in log(sSFR/yrs$^{-1}$) to define an active/passive galaxy.
}
\label{tabla_FG_sSFR}
\end{table}

\begin{figure}
\begin{center}
\leavevmode \epsfysize=8.6cm \epsfbox{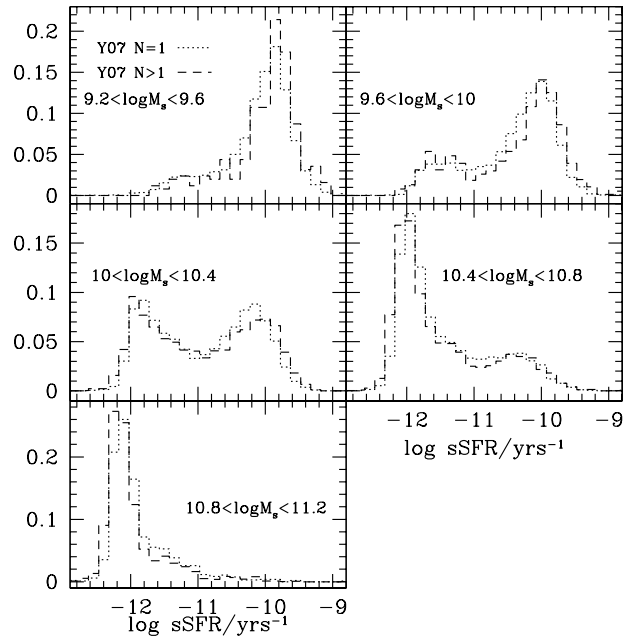}
\caption 
{Histograms of sSFR 
for field ($N$ = 1; dotted lines) and group (N $>$ 1; dashed lines) central galaxies for different stellar mass bins. The stellar mass \ms\ is in units of $\mstar$.
Each histogram was normalized by the total number of selected objects, so that
the sum of each distribution is equal to unity.
}
\label{FG_sSFR}
\end{center}
\end{figure}

Fig. \ref{FG_sSFR} shows the same as Fig. \ref{FG_gi} but for the sSFR.
Analogously to the color distributions, there are not 
significant
differences in this property between group and field 
central galaxies
(the chi-square test confirms a very high probability, $P > 0.998$, that both normalized distributions are similar in each panel).
In Table \ref{tabla_FG_sSFR} we report the corresponding fractions of active central galaxies, $f_{active}$ 
(the complement is the fraction of passive centrals, $f_{passive}=1-f_{active}$) in \ms\ bins as shown in Fig. \ref{FG_sSFR}.
The bimodality in sSFR is more evident for intermediate masses (middle panels).
In the mass range 10 $<$ log($M_{s}/\mstar$) $<$ 10.4, group central galaxies show two strong peaks 
(the active and passive populations), where the peak of active galaxies is slightly lower than that of 
passive 
objects. In the case of field central galaxies, the active and passive populations
exhibit similar peaks. However, the fractions of 
active group centrals and active field centrals are very similar (52\% and 53\%, respectively).  
For high-mass galaxies (bottom panel), 
most of the galaxies are passive.

\begin{figure}
\begin{center}
\leavevmode \epsfysize=8.6cm \epsfbox{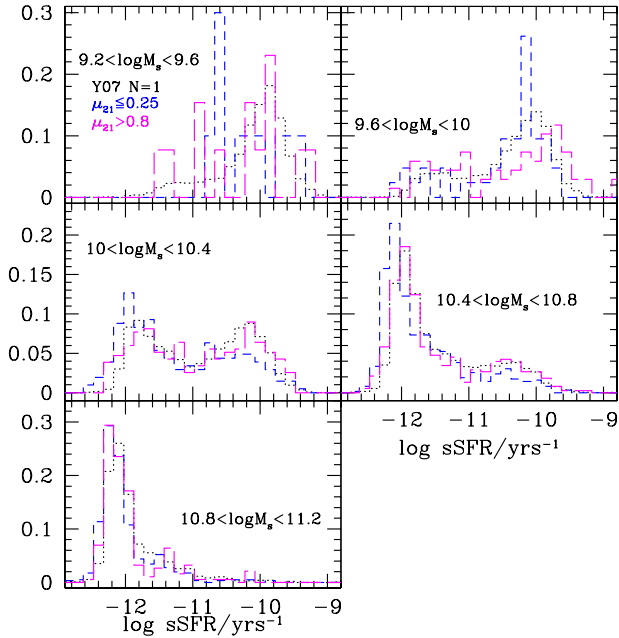}
\caption 
{Same as Fig. \ref{FG_sSFR}, but the group central galaxies (N $>$ 1) are split according to the stellar mass ratio, $\mu_{21}$, between the most massive satellite and the group central galaxy. 
Blue dashed and magenta long-dashed lines correspond to ratios lower than 0.25 and higher than 0.8, respectively.
Each histogram was normalized by the total number of selected objects, so that
the sum of each distribution is equal to unity.
The fraction of passive/active central galaxies with N $>$ 1 depends on the $\mu_{21}$ value. Central galaxies with $\mu_{21} >$ 0.8 (i.e., their most massive satellite has a comparable stellar mass) are slightly more active.}
\label{gap_sSFR}
\end{center}
\end{figure}

We further explore the possibility of some systematic variation in the color and sSFR for the group centrals 
as a function of the mass ratio with respect to the most massive satellite. The presence of a massive neighbor satellite
galaxy is expected to increase the probability of triggering star formation activity in the central galaxy. 
We select those group centrals ($N>1$) according to the stellar mass ratio
between the most massive satellite and the central galaxy
(i.e., the mass ratio between the second most massive and the most massive galaxy within the halo, $\mu_{21}$).
We only choose systems where the most massive satellite is complete in stellar mass by means of eq. (\ref{eq_Mslim}).
The distribution of $\mu_{21}$ for 
the selected sample has a broad
maximum at $\approx 0.25$, and correlates weakly  with the stellar mass in the sense
that the smaller the value of $\mu_{21}$, the larger the \ms\ of the group central galaxy
(for extensive results on this kind of dependence
on \ms\ and \mh\ see \citealp{Yang+2008}; \citealp{Rodriguez-Puebla+2013a}. For
similar results but with dependence on luminosity see \citealp{More2012,Hearin+2013}).

We note that the $g-i$ color distributions in different mass bins do not differ between those 
group central galaxies with $\mu_{21} > 0.8$ and $\mu_{21}\leq0.25$. 
In contrast, as can be seen in Fig. \ref{gap_sSFR}, small differences are found in the distributions of sSFR for group centrals in terms of $\mu_{21}$, in particular in the mass interval 9.6 $<$ log($M_{s}/\mstar$) $<$ 10.4. 
The group central galaxies ($N>1$) are split into two subsamples:
$\mu_{21}\leq0.25$ (blue dashed lines) and $\mu_{21}>0.8$ (magenta long-dashed lines). 
In general, from the figure we see that the overall fraction of active/passive central galaxies with N $>$ 1 depends on the value of $\mu_{21}$.
Active group central galaxies with $\mu_{21}>0.8$ have slightly higher sSFR values than those with  
$\mu_{21}\leq$ 0.25 since they peak at higher sSFR values.
The fraction of active centrals with $N>1$ is around 51\% for the mass ratio $\mu_{21}>$ 0.8, while for the mass ratio
$\mu_{21}\leq$ 0.25 this fraction is
38\% 
at the mass bin 10.0 $<$ log($M_{s}/\mstar$) $<$ 10.4 (left-middle panel).  
This suggests 
that central galaxies with high-mass ratios $\mu_{21}$ might have enhanced their 
star formation activity, most likely induced 
by their massive satellites.

It is worth noting that in the case of sSFR, the bimodality is well established for all central galaxies at the 
$\sim 1-5 \times 10^{10} \mstar$ masses, while the $g -i$ color distributions in most of the
mass bins are hardly bimodal (an incipient bimodality is seen for \ms $<$ $10^{10} \mstar$).
For intermediate-mass central galaxies, a fraction of $70-80\%$ are red and $30-50\%$ are active, thus
one expects a non-negligible fraction of red but currently star forming 
central galaxies. In Section \ref{reasons_quenching} we will discuss 
this result.

\subsection{Bright field and isolated central galaxies}
\label{results_F_I}

\begin{table}
  \centering
\caption{Relative fractions of blue bright field and blue isolated central galaxies
}
\begin{tabular}{c c c c c}
\hline
\hline
log $M_s$    &  $g-i$ cut   & $f_{blue}$ bright field & $f_{blue}$ isolated \\   
\hline
9.2--9.7     & 0.96     &  0.90      &  0.91 \\
9.7--10.2    & 1.04     &  0.70      &  0.76 \\
10.2--10.7   & 1.12     &  0.48      &  0.55 \\
10.7--11.2   & 1.20     &  0.34      &  0.35 \\                      
\hline
\end{tabular}

\tablecomments{\ms\ bins as shown in Fig. \ref{IF_gi}. The second column indicates the cut in $g-i$ to define a blue/red galaxy.
}
\label{tabla_IF_gi}
\end{table}

\begin{figure}
\begin{center}
\leavevmode \epsfysize=8.6cm \epsfbox{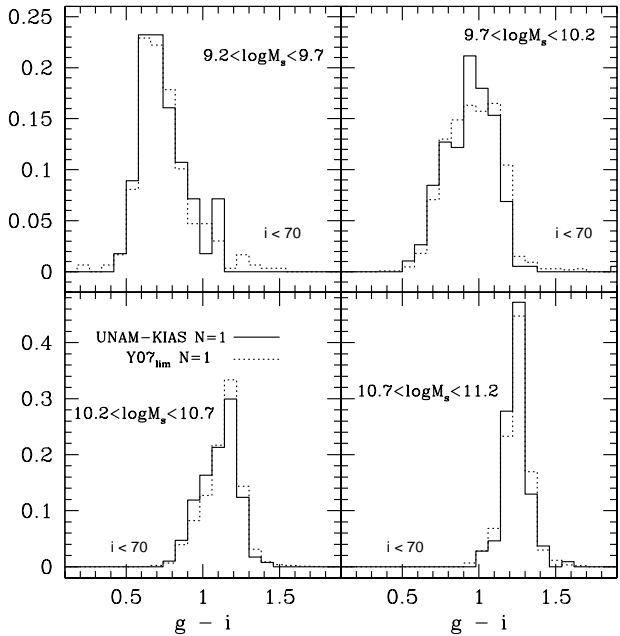}
\caption {
Histograms of $g - i$ color for bright field (dotted lines) and isolated (solid lines) central galaxies 
for different stellar mass bins (in units of $\mstar$). 
We do not include objects with high inclination (angles over 70 degrees).
Each histogram was normalized by the total number of selected objects, so that
the sum of each distribution is equal to unity.
}
\label{IF_gi}
\end{center}
\end{figure}

We proceed now to compare our samples of (bright) field and very isolated central galaxies. 
Recall that the latter is 
a subsample of the former.
Fig. \ref{IF_gi} shows $g - i$ color histograms
for bright field (dotted lines) 
and isolated (solid lines) centrals in different stellar mass bins. 
The distributions are roughly similar in general. The chi-square test in this case shows probabilities
above $P = 0.907$ that both normalized distributions are similar in each panel.
Table \ref{tabla_IF_gi} reports the fractions of blue central galaxies in bins of \ms\ corresponding
to the distributions shown in Fig. \ref{IF_gi}. 
These fractions are close for both samples; if any, the isolated centrals show a very marginal excess of 
blue galaxies with respect to the overall sample of bright field centrals, in particular at masses
9.7 $<$ log($M_{s}/\mstar$) $<$ 10.7, where the differences are of 6--7\%. 
In the relatively low-mass bin, 9.7 $<$ log($M_{s}/\mstar$) $<$ 10.2 (right-top panel), while bright field centrals exhibit a 
slight bimodality with a red peak at $g - i$ = 1.1, the distribution of isolated galaxies is narrower with a strong peak at $g - i$ = 0.9.  
Both populations of $N=1$ central galaxies are mainly red in the high-mass regime with a strong peak at $g-i \sim 1.3$.

\begin{table}
  \centering
\caption{Relative fractions of active bright field and active isolated centrals 
}
\begin{tabular}{c c c c c}
\hline
\hline
log $M_s$    &  sSFR cut  & $f_{active}$ bright field  & $f_{active}$ isolated\\   
\hline
9.2--9.7     & -10.5     &  0.83      &  0.80 \\
9.7--10.2    & -10.8     &  0.70      &  0.67 \\
10.2--10.7   & -11.2     &  0.43      &  0.50 \\
10.7--11.2   & -11.5     &  0.15      &  0.13 \\             
\hline
\end{tabular}

\tablecomments{\ms\ bins
as shown in Fig. \ref{IF_sSFR}. The second column is 
the cut in log(sSFR/yrs$^{-1}$) to define an active/passive galaxy.
}
\label{tabla_IF_sSFR}
\end{table}

\begin{figure}
\begin{center}
\leavevmode \epsfysize=8.6cm \epsfbox{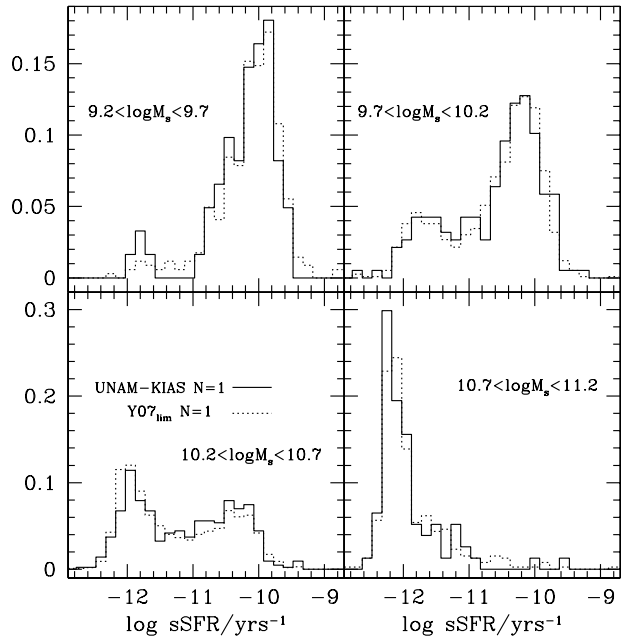}
\caption {
Histograms of specific star formation rate for bright field (dotted lines) and isolated (solid lines) central galaxies 
for different stellar mass bins (in units of $\mstar$). 
Each histogram was normalized by the total number of selected objects, so that
the sum of each distribution is equal to unity.}
\label{IF_sSFR}
\end{center}
\end{figure}

In Fig. \ref{IF_sSFR}, the sSFRs distributions for the bright field and isolated centrals are plotted in the same \ms\ bins 
as in Fig. \ref{IF_gi}. Again the distributions are similar in general. The chi-square test confirms this  
with probabilities above $P = 0.945$ that both normalized distributions are similar in each panel.
Table \ref{tabla_IF_sSFR} reports the fractions of active central galaxies 
in bins of \ms\ corresponding to the distributions shown in Fig. \ref{IF_sSFR}. 
For intermediate masses (left-bottom panel),  
the fraction of active galaxies is higher in the isolated sample by 7\% than in the bright field sample. 
The most massive regime (right-bottom panel) is dominated by 
passive galaxies in both samples.

\begin{figure}
\begin{center}
\leavevmode \epsfysize=7cm 
\epsfbox{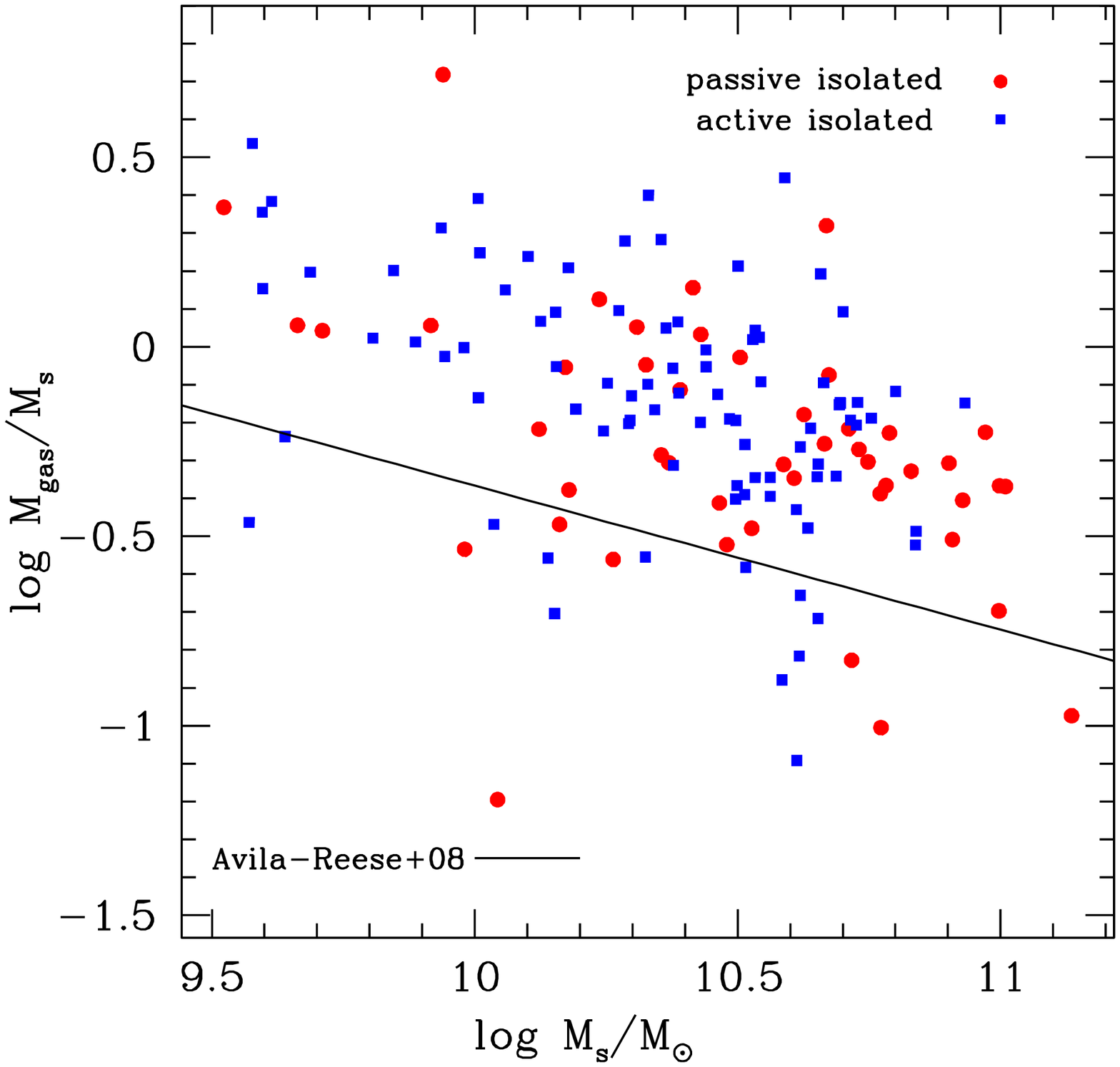}
\epsfysize=7cm 
\epsfbox{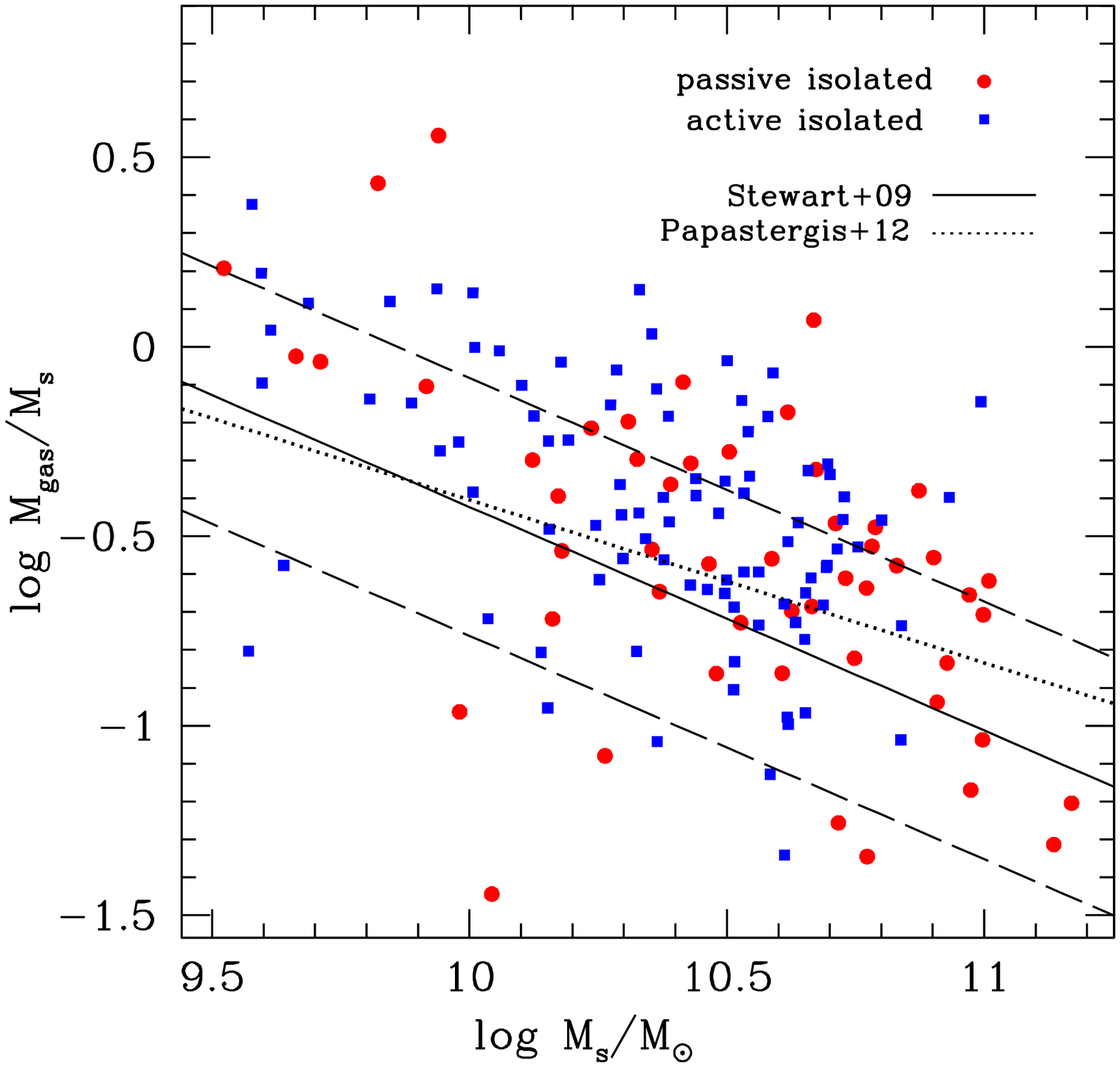}  
\caption{
Gas-to-stellar mass ratio, $M_{gas}/\ms$, 
as a function of \ms\ (here $h=0.7$)
for isolated central galaxies. They are split into two populations according to their sSFR; passive (red solid circles) and active (blue solid squares).
Top panel: We calculate $M_{gas}$ using the approach by \citet{Avila-Reese+2008} 
that takes into account helium, metals, and the mass of molecular hydrogen that depends on the morphology $T$. 
The solid line is the fit to their data.
Bottom panel: $M_{gas}$ is calculated using a correction for helium and metals, but  
mass of molecular hydrogen is not taken into account to be consistent with the equation (1) of Stewart et al. (2009, solid line. The dashed lines correspond to their reported 1$\sigma$ scatter).
In addition, the dotted line is the fit of \citet{Papastergis+2012} for observational estimates using H I mass.
Both panels show that low-mass isolated galaxies have in general higher 
$M_{gas}/\ms$ ratios
than the average relations.
}
\label{fgas_plot}
\end{center}
\end{figure}

\subsubsection{Gas-to-stellar mass ratios}

Fig. \ref{fgas_plot} shows the gas-to-stellar mass ratio, $M_{gas}/\ms$, for isolated galaxies with available information on H I mass (see Section \ref{secMs}).
They are split into two populations according to their sSFR; passive (red solid circles) and active (blue solid squares).
The top panel shows the gas mass $M_{gas}$ computed with the corrections for helium, metals, and H$_2$ mass (which depends on the morphological type $T$). 
Note that
most of the galaxies with available H I information plotted in Fig. \ref{fgas_plot} have late-type morphologies ($T>0$).
The solid line in the panel is the fit 
of \citet{Avila-Reese+2008} 
to a sample of local disk galaxies. In the case of elliptical galaxies, the $M_{gas}/\ms$ ratios are much smaller. 
Isolated centrals have on average higher $M_{gas}/\ms$ 
ratios compared to 
normal galaxies, for instance, by $\sim0.4$ dex at $\ms\approx 5\times10^{9} M _{\odot}$. 
The bottom panel shows the same as the top panel, but the gas mass is corrected only for helium and metals.
The solid and dotted lines correspond to the fits of \citet{Stewart+2009} and \citet{Papastergis+2012} for compilations 
of observational data of disk galaxies, respectively. The dashed lines show the 1$\sigma$ scatter 
according to \citet{Stewart+2009}.  As before, we see that isolated centrals, especially the lower-mass ones, 
have higher $M_{gas}/\ms$ ratios than 
normal disk galaxies, although the difference is smaller compared to the observed one in the top panel.

Isolated central galaxies have on average higher gas-to-stellar mass ratios than disk galaxies located in all environments, specially
at lower masses.  However, as shown in Section \ref{results_F_I}, the sSFR's of isolated galaxies at 
different masses do not differ significantly from those of bright field galaxies. This implies that isolated galaxies
have larger gas reservoirs than galaxies in average environments but with similar levels of star formation activity. 
On the other hand, as can be seen in Fig. \ref{fgas_plot}, the gas-to-stellar mass ratio of isolated centrals is 
somewhat independent if they are classified as active or passive star forming galaxies.
A more detailed object-by-object analysis is necessary  
for confirming these results obtained here at a statistical level.

\section{Galaxy--halo mass connection}
\label{secMh_conn}

\begin{figure}
\begin{center}
\leavevmode \epsfysize=7cm \epsfbox{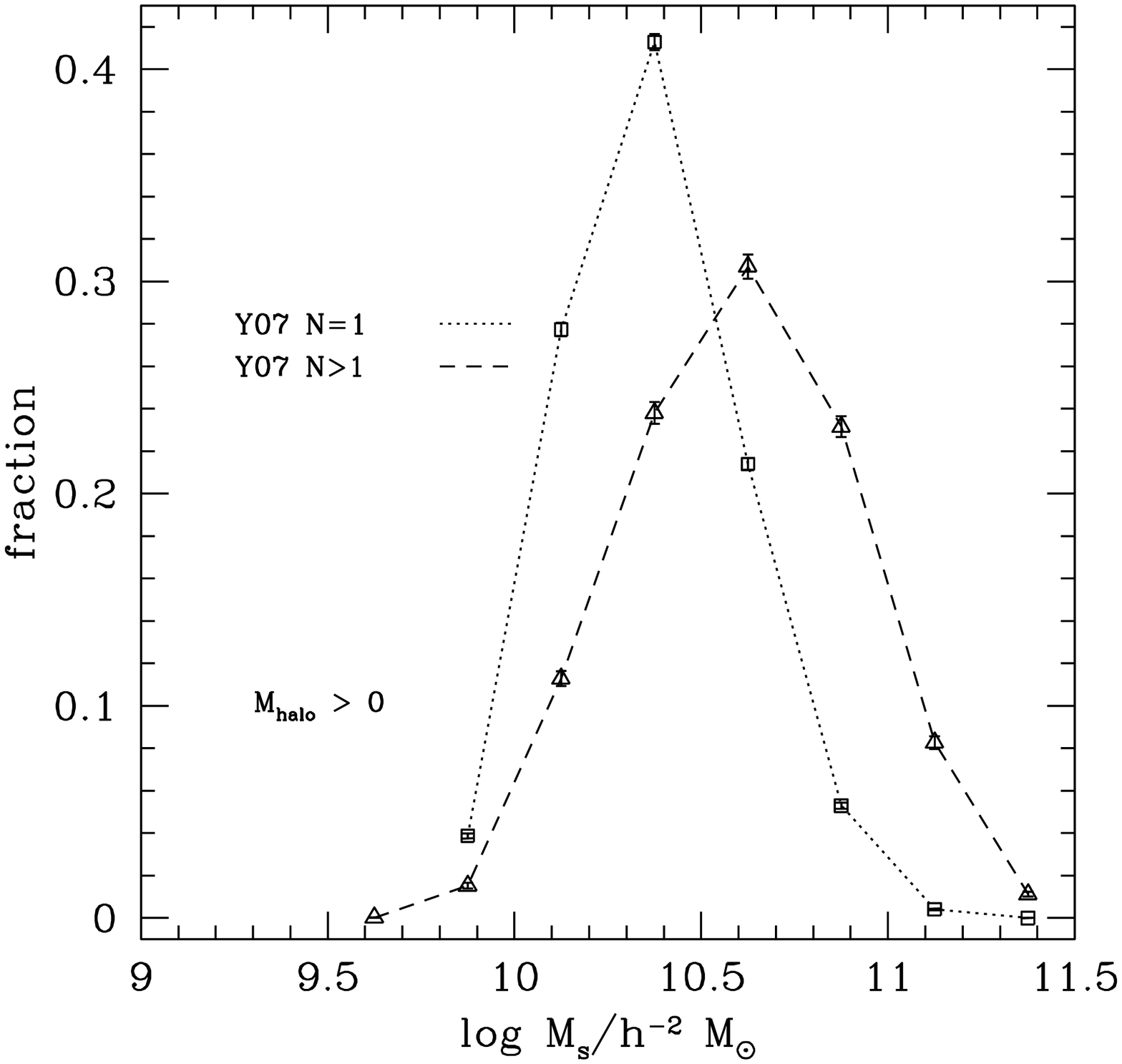} 
\leavevmode \epsfysize=7cm \epsfbox{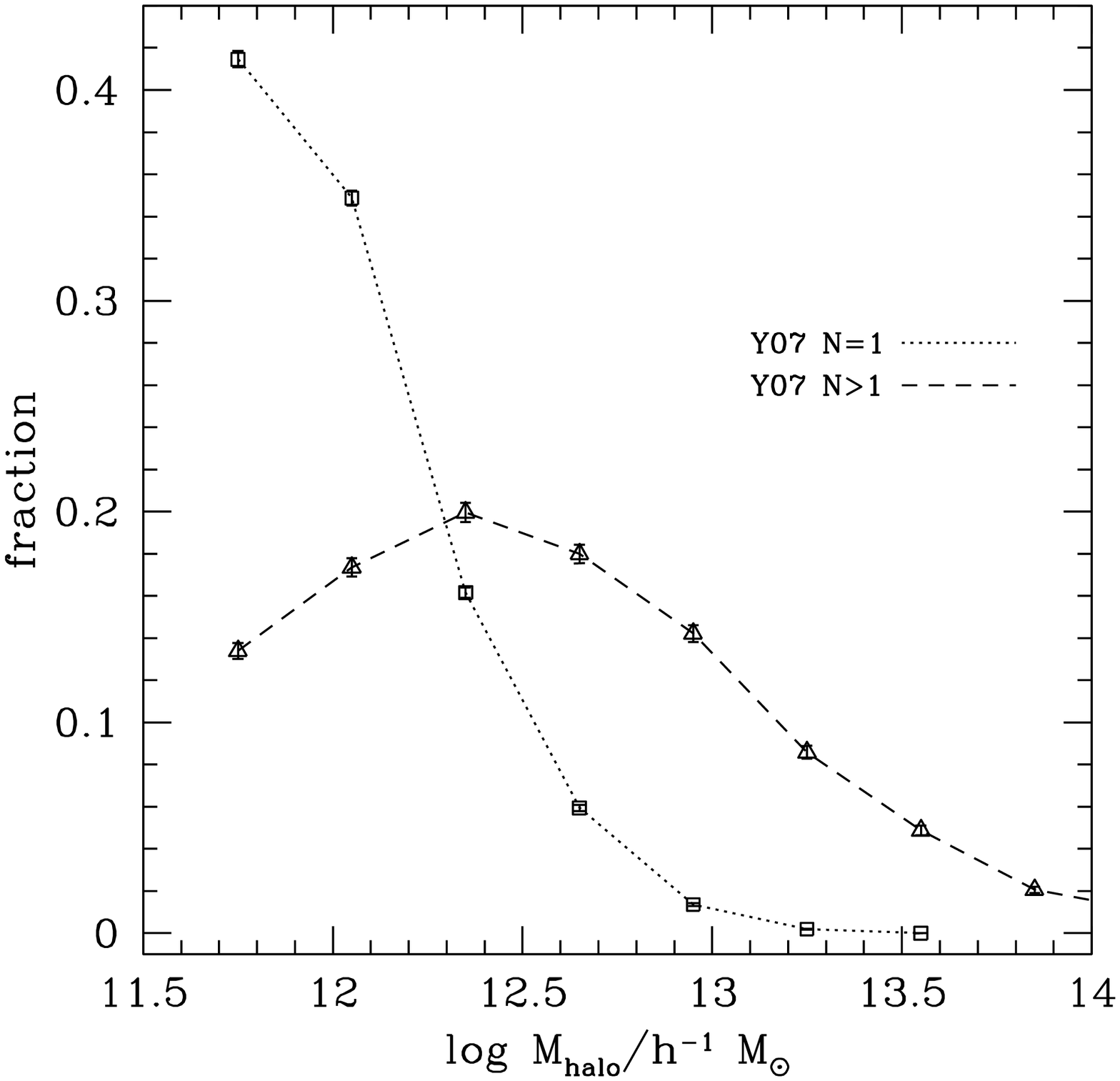} 
\caption 
{Top: Same as top panel of Fig. \ref{frac_Ms} but for central galaxies whose host halo mass is
inferred by Y07. Bottom: Relative fraction as a function of host halo mass. 
At high halo masses, log(\mh/$\mhalo$) $>$ 12.5, the fraction of group centrals dominates with respect to that of field central galaxies.
}
\label{frac_Ms-Mh}
\end{center}
\end{figure}

In this Section we use the available halo mass information from Y07 for our samples of central galaxies
(see details in Section \ref{secMh}). The top panel of Fig. \ref{frac_Ms-Mh} is the same as the top panel of Fig. \ref{frac_Ms}, but for central galaxies whose host halo mass is estimated by Y07. Since the determination of \mh\ for low-mass
galaxies is difficult, there are fewer low-mass galaxies (\ms $<$ $10^{10} \mstar$) than in 
Fig. \ref{frac_Ms} (see Table \ref{tabla_samples}). 
However, the trend at high-masses (\ms $>$ $10^{10.5} \mstar$) 
is still present; the relative fraction of group central galaxies is higher than that of field central objects. 
The chi-square test shows a very low probability of $P < 10^{-4}$ that both normalized distributions are similar.

The bottom panel shows the same relative fraction, but as a function of halo mass. 
The shape of the distributions among both samples are very different ($P < 10^{-7}$ that they are equal). Field ($N=1$) central galaxies 
have a steep decrease in the relative fraction as \mh\ increases. The probability distribution to 
find a field central object in halos of \mh $>$ $10^{12.5} \mhalo$ is low. 
In the case of central galaxies in halos which host satellites (group centrals), they have a wide distribution in halo 
masses, with a long tail in the massive regime.  The probability distribution to find a group central galaxy in halos of 
\mh $>$ $10^{12.5} \mhalo$ is relatively high in the local universe.  
It is well known that the halo mass correlates with the group/cluster richness \citep[see e.g.,][and the references therein]{Reyes+2008}.

We note that the relative fraction distribution of isolated central galaxies is similar  
to the one of bright field central galaxies, though the halo masses of the former are slightly smaller 
than those of the latter.
\citet{Niemi+2010} performed a study of isolated (elliptical) central galaxies using a \lcdm-based catalog 
of simulated galaxies. They found that these galaxies reside in dark matter halos with masses lower than 
$\sim$2 $\times$ 10$^{13}$ $\mhalo$, which is consistent with our results.

\subsection{The \ms--\mh\ relation}
\label{ref_MsMh}

In the last years, by means of statistical approaches and some direct measurements, the \ms--\mh\ 
relation at $z=0$ and at higher redshifts has been inferred and used as a key constraint for models and simulations of galaxy evolution
\citep[e.g.,][]{Mandelbaum+2006,Conroy+2009,Guo+2010,Behroozi+2010,Firmani+2010,Moster+2010,More+2011,Yang+2012,Behroozi+2013,
Moster+2013, Rodriguez-Puebla+2013a}. This relation sheds light on the efficiency of galaxy (stellar) formation as a function of 
halo mass, and its scatter constrains the role of internal and halo evolutionary effects and of environment on 
the final stellar mass of galaxies. Since the effects of  environment in isolated galaxies are in 
principle minimized, then the \ms--\mh\ relation of isolated galaxies should be associated mainly to 
intrinsic evolutionary processes. Does this relation differ from the one of all the galaxies?
Does its scatter correlate with the main galaxy properties?

Here, we compare \ms\ vs. \mh\ for our samples of central galaxies separated into groups according
to their properties. As a reference, we also compare our results  
with the fit to the \ms--\mh\ relation 
found in \citet{Yang+2009} for the Y07 data of central galaxies including the 1$\sigma$ scatter 
(the sample out to $z$ = 0.2 has been used by these authors). The reported constant scatter in \ms\ around 
this relation is 0.173 dex, though as these authors discuss, at lower masses it seems to be smaller. 
For consistency, we use here
the color separation into blue and red galaxies given by the same authors. They 
use the  $^{0.1}(g-r)$ color ($K$-correction and evolution-correction at $z$ = 0.1)
with Petrosian magnitudes, which are measured within a circular aperture defined 
by the shape of the light profile.
The $^{0.1}(g-r)$ color-magnitude criterion given in eq. (1) of Yang et al. (2008) is applied
to select blue and red galaxies.  For selecting passive and active galaxies, our eq. (\ref{eq_sSFR}) 
is used. Furthermore, for the isolated centrals, we separate the sample into early- and late-type galaxies
by the criterion $T$ $<$ 0 and $T$ $>$ 0, respectively.

\begin{figure}
\begin{center}
\leavevmode \epsfysize=8.6cm \epsfbox{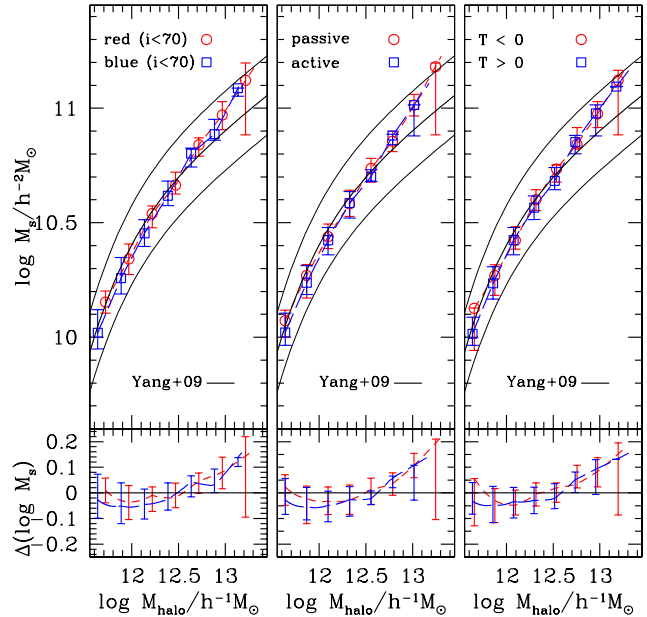} 
\caption {Stellar mass - halo mass 
relation for isolated central galaxies. They are split according to the color, 
sSFR and morphology (left, middle and right panels, respectively). The median distribution for red, passive and early-type galaxies is shown in red circles (red dashed lines), and that for blue, active and late-type galaxies is shown in blue squares (blue long-dashed lines).
Error bars correspond to the 16 and 84 percentiles.
\ms--\mh\ relations are very similar among these populations within the errors. 
The solid lines show the fit by Yang et al. (2009) to their data including the 1$\sigma$ scatter.
The lower box in each panel shows the residual in stellar mass, 
calculated as the difference between the median of each population and the fit by Yang et al. (2009), as a function of halo mass. Error bar corresponds to the dispersion of the median.
}
\label{I_MsMh}
\end{center}
\end{figure}

In the main panels of Fig. \ref{I_MsMh} we plot the \ms--\mh\ relation for our very isolated galaxy sample,
splitting it according to the color, sSFR, and morphological type $T$ (left, middle and right panels, respectively). 
The median distribution for red, passive, and early-type galaxies is shown in red circles (red dashed lines), 
and that for blue, active and late-type galaxies is shown in blue squares (blue long-dashed lines).
Error bars correspond to the 16 and 84 percentiles. As can be seen from each panel, the \ms--\mh\ relations 
are similar within the errors. This means that the \ms--\mh\ relation of isolated galaxies 
is nearly independent of color, sSFR, and morphology. 
This result is obtained by using the halo mass based on the characteristic stellar mass. 
If we use \mh\ based on luminosity ranking, a very
slight segregation by color in the \ms--\mh\ relation should appear but it is
entirely due to the mass-to-light ratio dependence on color, and not due to
something intrinsically related to the halo mass.

\begin{figure}
\begin{center}
\leavevmode \epsfysize=7.cm \epsfbox{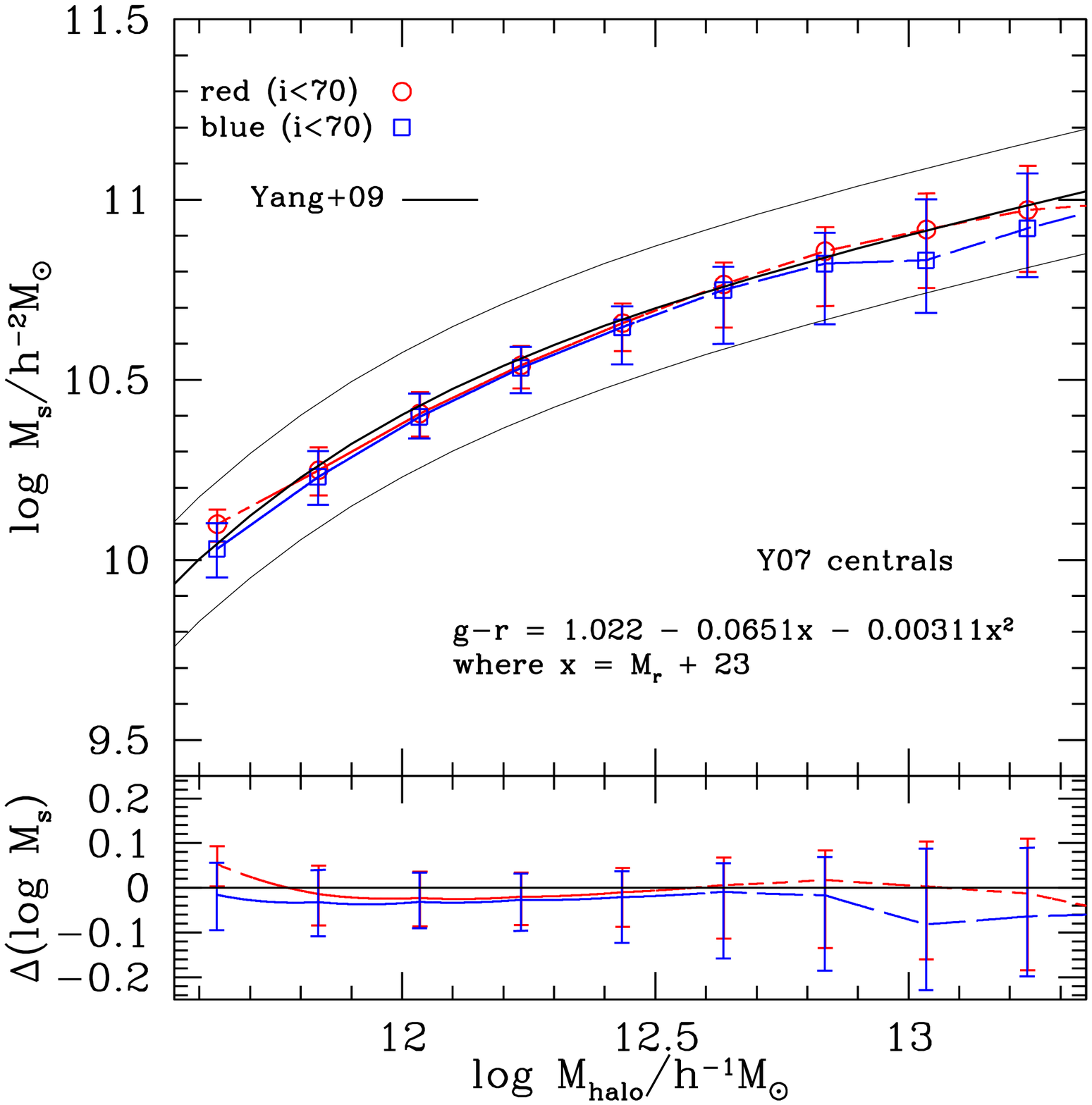} 
\epsfysize=7.cm \epsfbox{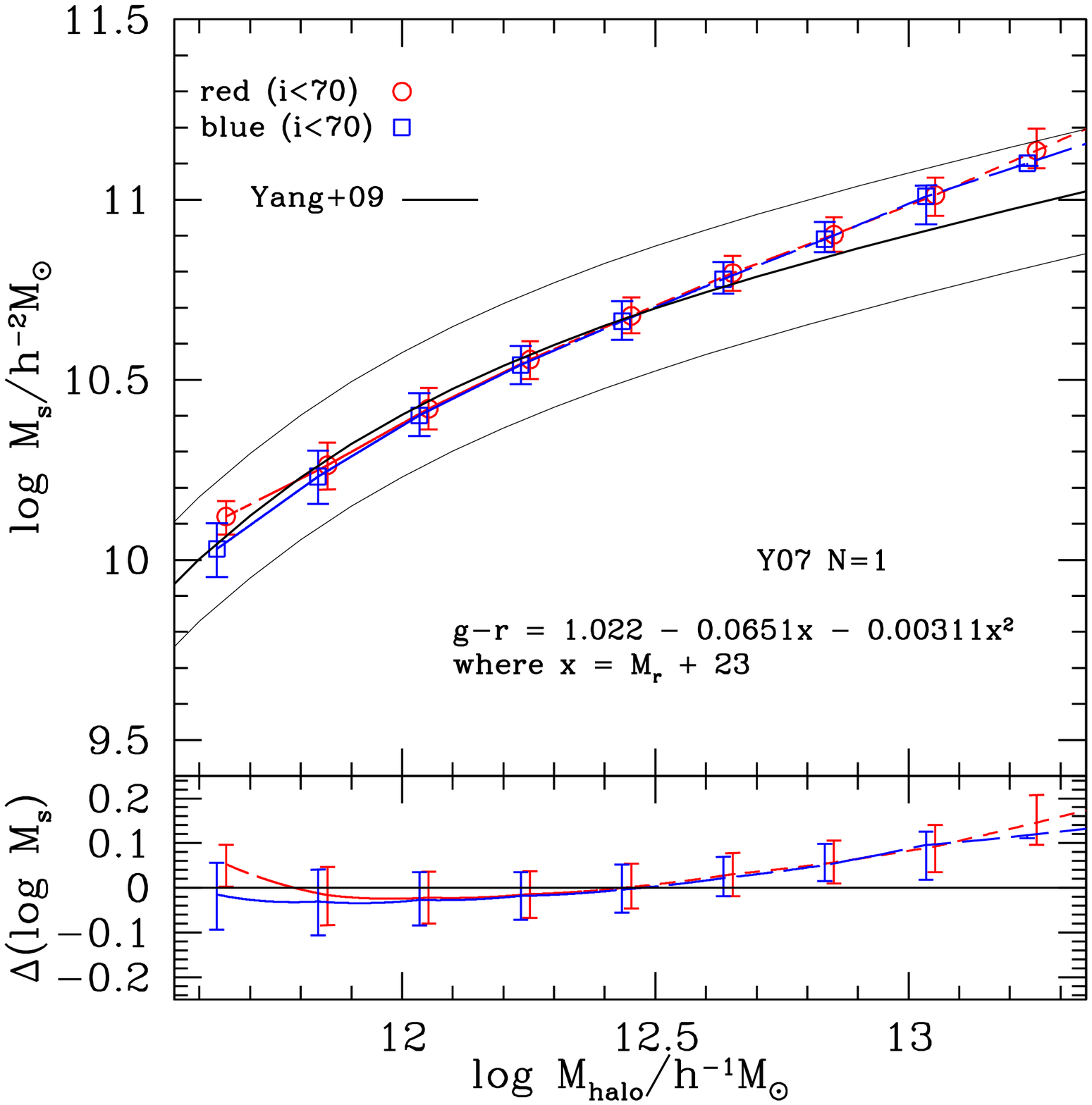} 
\caption {\ms--\mh\ relation for all (top panel) and field (bottom panel) central galaxies split by color. 
The median distribution for red galaxies is shown in red circles (red dashed lines), and that for blue galaxies is shown in blue squares (blue long-dashed lines).
Error bars correspond to the 16 and 84 percentiles.
The solid lines show the fit by Yang et al. (2009) to their data including the 1$\sigma$ scatter. Note that their sample reaches $z$ = 0.2, whereas our redshift limit is $z$ = 0.08.
The lower box in both panels shows the residual in stellar mass, 
calculated as the difference between the median of each population and the fit of Yang et al. (2009), as a function of halo mass. Error bars correspond to the dispersion of the medians. 
Field ($N$ = 1) central galaxies show a higher residual 
compared to that of all ($N$ $\geq$ 1) central objects at high masses,
which is independent of the galaxy property, e.g., color.
This different behavior of the  \ms--\mh\ relation is probably due to lack of systems without satellite galaxies in massive 
halos when assuming a one-to-one relation between total stellar mass and halo mass.
}
\label{all_MsMh}
\end{center}
\end{figure}

We compare the results in Fig. \ref{I_MsMh} with the 
fit to the \ms--\mh\ relation and its scatter found in \citet{Yang+2009}, which are shown as solid lines repeated in the main box of each panel.
The scatter for our sample of isolated galaxies is clearly lower than 0.173 dex at all masses. 
The implications of this result will be discussed in Section \ref{scatters}.
In addition, the lower box in each panel of Fig. \ref{I_MsMh} shows the residual in stellar mass, 
calculated as the difference between the median of each population and the fit by \citet{Yang+2009}, 
as a function of halo mass. Error bar corresponds to the dispersion of the median. The residuals show no 
differences among isolated galaxies of different properties (red/blue, passive/active or early-/late-type) within 
the errors, as already mentioned above.

According to Fig. \ref{I_MsMh}, the \ms--\mh\ relation of very isolated galaxies is similar to that one of the overall
sample of central galaxies in Y07 up to $\mh\sim 10^{12.5} \mhalo$.
For larger masses, there is a systematic deviation of the isolated galaxies towards larger stellar masses as \mh\ is
larger with respect to the overall sample of Y07 (out to $z=0.2$). This means that at a given halo mass, the stellar masses of the isolated
centrals are larger than those of the centrals in general. 
However, this is an expected result 
for centrals 
without satellites (our field $N=1$ sample) in massive halos 
when using the method of Y07 to estimate the halo mass. Under the assumption of the one-to-one relation between total (characteristic) stellar mass and halo mass,
when only one galaxy 
inhabits a massive halo, 
i.e. the total stellar mass in the system is the stellar mass of this (central) galaxy,
its \ms\ is larger than the mean for that halo mass.
To the inverse, when more satellites are present 
in the massive halo, the stellar mass of the central galaxy is smaller.
In Fig. \ref{all_MsMh} we plot the \ms--\mh\ relations (split by color)
for our Y07-based sample (out to $z$ = 0.08) of all central galaxies (field and group; upper panel), and for the field 
centrals ($N=1$, lower panel) only.
It is clearly seen that the \ms--\mh\ relation at large masses deviates from the 
average for the sample of $N=1$ centrals, where the isolated ones are contained.
In fact, it is rare 
to have only one galaxy in massive halos; they are populated typically by many galaxies 
(groups and clusters of galaxies).
Instead, when centrals with $N$ $>$ 1 are included (they dominate in number at large masses), 
the median relation is very similar to the fit of \citet{Yang+2009}. 
Although the difference shown in Fig. \ref{all_MsMh} in the \ms--\mh\ relations for centrals in halos with $N=1$ and $N>1$ is 
mostly due to the method used to construct the group catalog,
this may be still a valid result. For example, \citet{Niemi+2010} found that, for a given halo mass, simulated isolated elliptical central galaxies (which on average correspond to systems with $N=1$) contain more stellar mass with respect to simulated ellipticals in systems which roughly correspond to centrals in halos with $N>1$.

We also see in Fig. \ref{all_MsMh} that the \ms--\mh\ relation of all the centrals or only the $N=1$ (field) centrals is also
the same for blue and red galaxies, as was the case for the isolated galaxies. Therefore, we conclude that 
\textit{the \ms--\mh\ relation is nearly independent of the color for all kinds of central galaxies (group, field and isolated).}

A caveat to be taken into account is that the halo masses we are using were estimated by 
applying the abundance 
matching of the total stellar (characteristic) and the halo cumulative mass functions. Strictly speaking, the abundance matching 
should be carried out separately for the blue and red stellar mass functions and the corresponding (unknown) 
halo mass functions of both populations. Under some assumptions, this exercise has been done in 
\citet{Rodriguez-Puebla+2011}. They have found that the \ms--\mh\ 
relations for blue and red central galaxies are indeed close, though with more notorious differences than 
those shown in Fig.  \ref{all_MsMh}.

Finally, according to the results shown in Figs. \ref{I_MsMh} and \ref{all_MsMh},  
the scatters of the \ms--\mh\  relations
for the field centrals (including the isolated ones) seem to be lower than for the group central galaxies.
However, 
we caution that the absolute values of the scatters 
are only lower limits due to the way the halo mass has been assigned in the Y07 catalog (see also \citealt{Yang+2008}). 
We discuss further this topic in Section \ref{scatters}.

\section{Discussion}
\label{reasons}

An exhaustive comparison among only central galaxies in some different environments
was carried out. We compared the stellar and halo mass distributions among field 
and group centrals, and also between very isolated and bright field central galaxies.
In addition, we compared their colors and sSFRs in the same stellar mass bins. 
It is important to note that unlike previous studies of this kind, in which galaxy
properties are analyzed as a function of their environment without 
distinction between centrals and satellites \citep[for a review see][]{Blanton+2009}, 
or even in those works 
that separate galaxies into centrals and satellites
to study the differences
among the properties of both populations \citep[e.g.,][]{Weinmann+2006}, our analysis
have focused exclusively on central galaxies
in different environments.
Thus, it is difficult to compare our results with previous works. Following, we will discuss
whether some of our results are consistent or not with previous studies of isolated galaxies
or galaxies in systems with one or more satellites.

One of the remarkable differences between central galaxies in different environment is the stellar mass distribution. 
Field centrals ($N=1$) dominate at the low-mass end, whereas group centrals ($N>1$) dominate at the high-mass end. 
In this context,
\citet{Trinh+2013} used the NYU-VAGC catalog (based on SDSS DR6) to identify central galaxies in groups with $N=1$ and $N=2$. They found a higher fraction of centrals in halos with $N=2$ with respect to centrals in halos with $N=1$ at high stellar masses, which is consistent with our results.

On the other hand, \citet{Trinh+2013} found a slight blue excess of 6\% for centrals in systems with $N=2$ relative to field centrals ($N=1$) with the same stellar mass distribution using a  sample with $M_r\le-19$. This excess is consistent with our results but
it is somewhat higher since
we detect a very small excess of blue group centrals with respect to field centrals, 
around 1$-$5\% (see Table \ref{tabla_FG_gi}). 
These authors defined red/blue galaxies using $g-r=0.68$. If we use their condition in the $g-r$ color in addition to the cut  $M_r\le-19$, the population of red galaxies vanishes at very low masses (top-left panel in Fig. \ref{FG_gi}).  
We do not find any important excess of blue group central galaxies 
in systems with $N=2$ (the differences are typically smaller than 4\%), except the interval 9.6 $<$ log($M_{s}/\mstar$) $<$ 10 that shows a blue excess around 7\% for centrals in systems with $N=2$ relative to field centrals. 
The latter is in rough agreement with their results.

When comparing the subsample of very isolated galaxies with the 
sample of bright field galaxies 
(both with the same apparent magnitude limit), 
we do not find strong differences in their sSFR distributions. 
In 
agreement with our results, 
\citet{Kreckel+2012} did not find strong 
differences in sSFR between 
low-mass SDSS galaxies (centrals and satellites) and
a sample of galaxies in voids (large-scale regions of extreme low-density 
environment where most of 
the isolated galaxies could reside) with the same stellar mass.
On the other hand, \citet{Rojas+2005} and \citet{vonBenda-Beckmann+2008} found that relatively bright galaxies in voids have slightly higher sSFR values than similar galaxies located in higher density environments. This is again in rough agreement with our results since our isolated central galaxies at intermediate masses (10.2 $<$ log($M_{s}/\mstar$) $<$ 10.7) show slightly higher sSFR values than bright field galaxies (see Table \ref{tabla_IF_sSFR}). Note that the control samples in this kind of works correspond in general to central and satellites galaxies, whereas our comparison of isolated central galaxies was done only with respect to bright central galaxies in halos with $N=1$, i.e., 
we do not include satellites galaxies nor central galaxies in halos hosting also satellites.

\subsection{Mass growth of central galaxies}
\label{disc_MassGrowth}

We have seen that the mass distribution of group and field galaxies is different; 55\% of the group centrals is found in the massive regime (stellar masses above the Milky Way), whereas this fraction is smaller than $\sim$20\% for the field centrals.
Group central galaxies live typically in higher-density environments. 
As can be seen in the bottom panel of Fig. \ref{frac_Ms-Mh}, most of these galaxies are located in massive host halos, 
which correspond to groups and clusters of galaxies. In this kind of environments, it is probable that the central
galaxy suffered several mergers in the past, thus attaining higher stellar masses than the field galaxies.
As the system (halo) grows up lately, the internal velocity dispersion increases in such a way that the merger probability
of the remaining/new members becomes very low and today it remains as a system of many ($N>1$) members
in virial equilibrium.

On the other hand, the mass distribution of very isolated central bright galaxies is close to that of field bright centrals 
(both in systems with $N=1$). 
\citet{Hirschmann+2013} 
used a semi-analytic galaxy formation model based on \lcdm\ cosmology and found that minor
mergers provide a larger contribution to the overall stellar mass assembly of isolated galaxies as compared to major mergers.
It is thought  that minor mergers do not modify the morphology of the main galaxy. We check the morphological type for the 
isolated central galaxies with \ms $\geq$ 10$^{10.5}$ $\mstar$ and find that those with $T$ $\geq$ 1 (late-type 
galaxies from Sa to Sm) dominate with 62\%, whereas those with $T$ $\leq$ -1 (early-type galaxies from 
S0 to ellipticals) are just 31\% of the sample at these masses (the remaining 7\% corresponds to 
intermediate morphologies with -1 $<$ $T$ $<$ 1). Thus, 
late-type morphologies are more common than early-type ones in
our sample of massive isolated galaxies.
This supports the idea that many of these galaxies increased their stellar masses
partially by means of minor mergers,
although there is also a fraction of early-type isolated galaxies that presumably grew through major mergers.
It is plausible that a similar situation occurs with the massive centrals from the bright field sample, i.e., 
a dominant population of late-type galaxies at high stellar masses inhabiting relatively low-mass halos.

We conclude that centrals formed in halos rich in members ($N>1$) are more probable to 
grow more in mass than those in halos with only one galaxy ($N=1$, including the very isolated galaxies), 
probably due to more mergers. On the other hand, for massive centrals, a galaxy
in a halo without satellites 
seems to have a smaller host halo mass than a galaxy of the same stellar mass but 
inhabiting a halo with many members. This may evidence that 
host halos of $N=1$ or $N>1$ centrals also have different
growth (merger) histories. 
All this suggests that the stellar and halo merging of central galaxies within halos with many 
members ($N>1$) should be different than those in halos without satellites ($N=1$). 
However, in Section 3 we showed that, for a given stellar mass, there are no remarkable differences 
in properties such as color and sSFR among the central galaxies. 
Therefore, it seems that \textit{the main galaxy properties and the evolution itself
of central galaxies is dominated by internal processes rather than by the merger histories and environmental effects.}
We will study elsewhere the growth of central galaxies of equal stellar mass but with diverse halo assembly histories using 
numerical simulations.

\subsection{Quenching of central galaxies}
\label{reasons_quenching}

In Section \ref{secObs} we presented observational results regarding the color and sSFR 
of the central galaxies. About 70--80\%  of field and group centrals at intermediate masses are red 
but there is a 30$-$50\% of them with ongoing star formation.  We have checked that
$\sim$20\% of the field galaxies with $10^{10} < M_{star}$ $<$ $10^{10.8} \mstar$ are indeed
classified as red but with high sSFR's. 
This fraction decreases to $\sim$15\% for group galaxies.
It is in general assumed that the quenching (i.e., when a galaxy passes from star forming to passive) occurs 
in blue or green galaxies before they become red (and dead) objects. The fraction of red but active 
centrals at intermediate masses in our sample certainly do not follow this assumption.
In addition, \citet{Mendel+2013} identified a sample of young passive (quenched) galaxies from SDSS and found they are predominantly early-type objects, which suggests that the quenching is accompanied by morphological transformations in galaxies of the local universe (bear in mind that their sample includes central and satellite galaxies).

It is a very hard task to identify the physical processes behind the quenching, specially for central galaxies. 
Empirical and semi-empirical studies have shown that on average very massive galaxies were in the phase of 
quenching at high redshifts but with cosmic time, smaller galaxies become
quenched (mass downsizing; e.g., \citealp{Bundy+2006,Drory+2008,Pozzetti+2010,Firmani+2010}).
According to these studies, the typical stellar mass at which the transition to the passive 
regime happens at $z\sim 1$ and $z\sim 0$ is $6-7\times 10^{10}$ \msun\ and 
$2-3\times 10^{10}$ \msun, respectively (see for instance, Fig. 8 in \citealp{Firmani+2010} and
more references therein).  
\citet{Peng+2010,Peng+2012} pointed out that different
physical mechanisms may mimic their simple 
relation for the quenching rate of central galaxies, which is proportional to the SFR of them.
On the other hand, \citet{Woo+2013} used the SDSS to show that the fraction of quenched central 
galaxies is strongly correlated with halo mass at fixed stellar mass. 
This suggests that the quenching of a central galaxy is due to halo mass-depending mechanisms 
that prevent the cooling of infalling gas
such as virial shock heating \citep{DB2006}. \citet{Yang+2013} support this picture by indicating that 
star formation in central galaxies is quenched once their halo masses reach a characteristic mass, 
which goes from $\sim$ 10$^{12.5}$ $\mhalo$ at $z$ $>$ 3.5 to $\sim$ 10$^{11.5}$ $\mhalo$ at $z$ = 0.
At larger halo masses, the gas is diluted and susceptible to AGN feedback.

The color transformation (from blue to red) should occur after quenching processes.
The recent star formation activity in some field and group red central 
galaxies is perhaps  due to interactions with other galaxies. It is known that star formation is moderately enhanced in galaxy 
pairs \citep[e.g,][]{Lambas+2003,Hernandez-Toledo+2005}, which extend out to projected separations of 
$\sim$150 kpc \citep[]{Patton+2013}, i.e., beyond the virial radius of low-mass halos.
Central massive galaxies at $z=0$ are the result of a tumultuous star formation history at $z\ge2$ but 
seem to follow a relatively quiet 
evolution since $z<1$. Most of the local central galaxies at intermediate stellar masses have old stellar populations 
(hence red colors) after quenching processes. We conclude that for the fraction of those that have signs of recent 
star formation activity, it is plausible that this happens due to 
interactions between galaxies (for field and, in lower degree, group centrals with perturbers).

\subsection{On the scatter of the \ms--\mh\ relation}
\label{scatters}

The semi-empirical \ms--\mh\ relation has received a lot of attention in the last years since it summarizes 
many aspects of the efficiency of galaxy formation as a function of halo mass (see the references 
in Section \ref{ref_MsMh}). 
According to these studies 
(see also Figs. \ref{I_MsMh} and \ref{all_MsMh}), at masses around 
$\mh\sim 10^{12}$ \msun, the efficiency of stellar mass assembly is maximal; at lower masses
it decreases, probably due to the stellar--driven (mainly Supernovae) negative feedback, as well as
at larger masses, due to the long gas cooling times and AGN--driven negative feedback.  All these
processes are strongly dependent on the gravitational potential determined mainly by the halo mass. 
Therefore, the halo mass seems to be the main driver of the stellar mass growth of galaxies. Are 
other physical and evolutionary factors relevant, e.g., the halo mass aggregation history, the gas
angular momentum,  the environment, etc.? 

Our analysis of only central galaxies presented in Section \ref{secMh_conn} 
has shown that:
(i) The \ms--\mh\ relation of very isolated bright centrals does not differ from that of all field ($N=1$) centrals.
(ii) There is not 
a significant
systematic difference in the \ms--\mh\ relation of central galaxies separated into blue 
and red (and into active and passive, and early- and late-types in the case of the isolated galaxies). 
(iii) The \ms--\mh\ relation of centrals with and without satellites ($N=1$ and $N>1$, respectively) are similar up to $\mh\sim 10^{13} \mhalo$; at larger masses, there are almost no centrals without satellites, but those few that are observed deviate systematically to larger stellar masses for their \mh\ 
than the $N>1$ centrals; 
the latter is actually a consequence of assuming a one-to-one relation between total stellar mass and halo mass
(see Section \ref{ref_MsMh}). 
Under the same assumption, the scatter of the \ms--\mh\ 
relation increases with halo mass for group central galaxies ($N>1$), being minimal for field/isolated galaxies ($N=1$).

The result
that the scatter
does not depend on color, sSFR or morphology may imply that all the physical and evolutionary processes that
coin these internal properties are not important for the efficiency of galaxy stellar mass growth. Indeed, as shown in Section 
\ref{results_F_I}, once the stellar mass is fixed, the color and sSFR do not differ significantly
between isolated and bright field centrals.

For the group central galaxies, neither the colors nor sSFRs  differ significantly from those of field centrals for a given
stellar mass. However, the scatter at large masses in the \ms--\mh\ relation 
seems to be higher for the group centrals.
Under the assumption of a one-to-one relation between total stellar mass ($M_{s,tot}$) and halo mass, 
for a given \mh\ (and therefore $M_{s,tot}$), as smaller is the contribution of satellites to $M_{s,tot}$, the larger is the \ms\ of the central galaxy. Thus, it is expected that the scatter around the \ms--\mh\ relation at group/cluster scales is partially due to this ``mass partition" effect.

It is common to assign a scatter of $\sim 0.17$ dex to the \ms--\mh\ relation based mainly on inferences
for group centrals, living in massive halos \citep[c.f. Y07;][]{Yang+2009}. However, as discussed above,
\textit{most of this scatter is dominated by conditions of the group establishment rather than by internal physical and evolutionary
processes of the central galaxies.}  
The latter processes should produce a scatter in the \ms--\mh\ relation of field/isolated ($N=1$) galaxies, which can be thought as the intrinsic component of the scatter around this relation in general. 
As mentioned above, Y07 calculated the group halo mass by assuming a one-to-one relation between the total stellar mass of the group and its halo mass (with no scatter in this relation), so that we cannot obtain conclusive results regarding the intrinsic scatter of the \ms--\mh\ relation by using this method.
It should be very relevant to measure this intrinsic scatter, 
by means of statistical semi-empirical approaches (e.g., \citealt{Rodriguez-Puebla+2013a}) and/or by using the next weak lensing surveys,
which 
puts strong constraints on models of galaxy evolution.

\section{Conclusions}
\label{secConcl}

We have studied in detail the properties and distributions of 
central galaxies by using a catalog of very isolated galaxies
\citep[UNAM-KIAS collaboration,][]{Hernandez-Toledo+2010}
and 
a large halo-based galaxy group catalog (Y07), both constructed from the SDSS. 
Central galaxies were initially divided into 
two environments: those with satellite(s) inside the halo virial radius ($N>1$, group 
centrals) and those without satellites 
($N=1$, field centrals).
From the latter, 
we select the subsample of very isolated central galaxies, i.e. galaxies without perturber neighbors up to very large radii (likely much beyond the virial radius) according to strict 3D-isolation criteria. In order to compare adequately this subsample with the one of field galaxies, a subsample from the latter with the same apparent magnitude limit of the isolated centrals
 ($m_r <  15. 2$) was selected (bright field centrals).
In order to attain similar completeness limits and homogeneity, the four samples 
are limited to the same redshift range $0.01 \le z \le 0.08$, they are complete in \ms,
and their galaxy properties were taken from the same analysis.
Our main results are as follows:

-- The stellar mass distributions of the field and group central galaxies are different (top panel of Fig. \ref{frac_Ms}). 
The relative fraction of field and group central galaxies dominate at $M_{s}/\mstar \sim$ 10$^{10.4}$ and 
10$^{10.6}$, respectively. The differences in the mass distributions explain why the blue/star forming regions in 
the color--\ms\ and sSFR--\ms\ diagrams are populated mainly by field central galaxies, whereas the group central 
galaxies are mostly biased to 
the red/passive regions (Fig. \ref{CMD_sSFR}).
In the case of the isolated centrals, 
they do exhibit the same occupation in these diagrams 
compared to bright field central galaxies since both 
have similar stellar mass distributions.

-- At parity of stellar mass (in the same \ms\ bins), the color distributions of the central galaxy samples are 
similar, 
specially between field and group centrals. 
If any, marginal differences 
arise at low-masses, where the relative fraction of blue group centrals is around 5\% higher than that 
of blue field centrals. 
On the other hand, the relative fraction of blue isolated central galaxies is up to barely
$\sim$7\% higher than that of the blue bright field centrals at intermediate masses.

-- The sSFR distributions of the central galaxy samples are also similar at parity of stellar mass, 
where bimodality is clearer than in the color distributions, specially for intermediate masses.
When we use the stellar mass ratio
between the most massive satellite and the central galaxy, $\mu_{21}$,
active group central galaxies with 
$\mu_{21}>0.8$ peak at higher sSFR values than those with
$\mu_{21}\leq$ 0.25 for stellar mass bins in the interval 9.6 $<$ log($M_{s}/\mstar$) $<$ 10.4.

-- The \ms--\mh\ relation of isolated galaxies (less affected by environment), shows no differences when they are separated by color, sSFR and morphological type $T$ (using the halo mass based on the total stellar mass content).
The same result is found for field centrals ($N=1$) and
group centrals ($N>1$) separated by color.

-- For isolated and field (both $N=1$) galaxies, the \ms--\mh\ relation steepens
at high halo masses with respect to group centrals; at these masses the probability to find a halo with a single member is actually low.
This deviation is explained as a condition of the group 
when assuming a one-to-one relation between total stellar mass and halo mass
(less members in a halo, more mass for the central galaxy) rather than by internal processes.
Under the same assumption,
the scatter around the \ms--\mh\
relation of group centrals ($N>1$), which are mostly massive and living in massive halos ($>10^{12.5}$ $\mhalo$), 
increases systematically with $M_h$ and it is likely higher than 
the scatter corresponding to
isolated and field central galaxies (both $N=1$). 
The absolute values of the scatters 
are only lower limits.

The general results from our study lead us to conclude that
centrals assembled in dense environments (groups and clusters of galaxies) tend to have a stellar mass 
distribution biased to large masses (\ms$> 10^{10.6} \mstar$), likely because they grew up significantly by
mergers in the past, but after the velocity dispersion increases as the large halo relaxes, the probability
of merging decreases and they remain as massive centrals surrounded by many satellite galaxies. In the case
of the field ($N=1$) centrals, most of them likely inhabit average environments in density 
(filaments/walls), where they have grown up less by mergers with the surrounding satellites, and may 
suffer the loss of some of them due to the tidal interactions with the larger structures in which their halos 
inhabit. Therefore, the mass distribution of field centrals tends to be biased to lower masses as compared 
with group centrals. 
In the case of the subsample of very (and bright) isolated galaxies, inhabiting probably in voids 
and the outskirts of walls and clusters, their mass distribution is similar to that of bright field galaxies, thus suggesting that both 
populations share comparable mass growth mechanisms.

The different mass distributions of the group and field centrals are behind the global differences
of these samples in what regards their colors and sSFRs values. However, at a given \ms\ bin, 
there are only minor differences among the distributions of these properties for them.
This implies that the mass growth of central galaxies is mostly driven by other factors and 
processes rather than the local environment and mergers.

The stellar mass growth efficiency of all the centrals (given by the \ms-to -\mh\ ratio) is 
tightly related to their halo masses. 
This suggests that the halo mass is 
the main 
driver of central galaxy \ms\ growth.  
Disentangling the 
intrinsic 
scatter 
in the \ms--\mh\ relation of  
central galaxies without satellites will certainly help us to understand the internal processes that give rise to the galaxy properties and their 
connection with the evolution of halos.

\section*{Acknowledgments}

We thank the referee for the thoughtful comments and suggestions that helped to improve the paper.
We thank Jos\'e A. V\'azquez-Mata for providing us the updated 
version of the UNAM-KIAS catalog. I.L. would like to thank Rut Salazar and 
Angel R. Calette for their technical support using CasJobs.
I.L. acknowledges support from the Postdoctoral Fellowship program of DGAPA-UNAM, Mexico. 
A.R-P. and V.A-R acknowledge CONACyT (ciencia b\'asica) grant 167332 
for a terminal graduate student fellowship and for partial support. 
H.M.H.T acknowledges support from DGAPA-PAPIIT IN-112912 grant.

\bibliographystyle{apj}
\bibliography{references}

\label{lastpage}

\end{document}